%% file: qsocgm.tex
\documentclass[iop]{emulateapj}

\newcommand{\lya}{Lyman-$\alpha$}
\newcommand{\cm}[1]{\rm~cm^{#1}}
\newcommand{\mnhi}{N_{\rm HI}}
\newcommand{\nhi}{$N_{\rm HI}$}
\newcommand{\rvir}{$R_{\rm vir}$}
\newcommand{\msun}{~\rm M_\odot}
\newcommand{\msol}{~\rm M_\odot}

\newcommand{\Jbk}{{\ensuremath{J_{\rm bk}}}}

\newcommand{\cc}{\ensuremath{{\rm cm^{-3}}}}

\shorttitle{Confronting Simulations of Optically Thick Gas in Massive Halos with Observations at $\lowercase{z}=2-3$}
\shortauthors{Fumagalli et al.}

\begin{document}

\title{Confronting Simulations of Optically Thick Gas in Massive Halos with Observations 
  at $\lowercase{z}=2-3$}

\author{Michele Fumagalli\altaffilmark{1,2,11}, 
Joseph F. Hennawi\altaffilmark{3}, 
J. Xavier Prochaska\altaffilmark{4,5}, 
Daniel Kasen\altaffilmark{6,7},
Avishai Dekel\altaffilmark{8},
Daniel Ceverino\altaffilmark{9}, 
Joel Primack\altaffilmark{10}} 

\altaffiltext{1}{Carnegie Observatories, 813 Santa Barbara Street, 
  Pasadena, CA 91101, USA. \email{mfumagalli@obs.carnegiescience.edu}}
\altaffiltext{2}{Department of Astrophysics, Princeton University, Princeton, NJ 08544-1001, USA.}
\altaffiltext{3}{Max-Planck-Institut f\"{u}r Astronomie, K\"{o}nigstuhl 17, D-69117 Heidelberg, Germany.}
\altaffiltext{4}{Department of Astronomy and Astrophysics, University of California,
  1156 High Street, Santa Cruz, CA 95064, USA.}
\altaffiltext{5}{UCO/Lick Observatory; University of California, 1156 High Street, Santa Cruz, CA 95064, USA.}
\altaffiltext{6}{Department of Physics, University of California, 366 LeConte, Berkeley, CA 94720, USA.} 
\altaffiltext{7}{Nuclear Science Division, Lawrence Berkeley National Laboratory, Berkeley, CA 94720, USA.}
\altaffiltext{8}{Racah Institute of Physics, The Hebrew University, Jerusalem 91904, Israel.} 
\altaffiltext{9}{Departamento de F\'isica T\'eorica, Universidad Aut\'onoma de Madrid, Cantoblanco, 28049 Madrid, Spain.}
\altaffiltext{10}{Department of Physics, University of California, 1156 High Street, Santa Cruz, CA 95064.}
\altaffiltext{11}{Hubble Fellow}

\begin{abstract}
Cosmological hydrodynamic simulations predict the physical state of
baryons in the circumgalactic medium (CGM), which can be directly
tested via quasar absorption line observations. We use high resolution
``zoom-in'' simulations of 21 galaxies to characterize the distribution
of neutral hydrogen around halos in the mass range $M_{\rm vir} \sim 2\times
10^{11} - 4\times 10^{12}~\rm M_\odot$ at $z\sim2$.
We find that both the mass fraction of cool ($T \le 3\times 10^4~\rm K$) 
gas and the covering fraction of optically-thick Lyman limit systems (LLSs) 
depend only weakly on halo mass, even around the critical value for the 
formation of stable virial shocks.  The covering fraction of LLSs 
interior to the virial radius varies between $f_{\rm c} \sim 0.05 - 0.2$, 
with significant scatter among halos. Our simulations of massive halos 
($M_{\rm vir} \ge 10^{12}~\rm M_\odot$) underpredict the covering fraction 
of optically-thick gas observed in the quasar CGM by
a large factor.  The reason for this discrepancy is unclear, but
several possibilities are discussed. In the lower mass halos 
($M_{\rm vir} \ge 5 \times 10^{11}~\rm M_\odot$) hosting star-forming 
galaxies, the predicted covering factor agrees with observations, 
however current samples of quasar-galaxy pairs are too small for a 
conclusive comparison.  To overcome this limitation, we propose a new 
observable:  the small-scale auto-correlation function of optically-thick 
absorbers detected in the foreground of close quasar pairs. We show that 
this new observable can constrain the underlying dark halos hosting LLSs 
at $z\sim 2-3$, as well as the characteristic size and covering factor of the 
CGM.
\end{abstract}


\section{Introduction}\label{intro}

Over the past several years, numerical simulations of galaxy formation have 
converged upon a paradigm for the accretion of gas into dark matter halos.  
One tenant of the model is that the majority of gas which travels to the central 
regions and contributes fuel for star-formation is cool, i.e. at a 
temperature $T \sim 10^4-10^5$\,K \citep{bir03,ker05,dek06,ocv08,ker09b,dek09,van11}.  
Importantly, the simulations reveal that this cool gas travels along relatively 
narrow (i.e.\ filamentary) structures often termed ``cold streams''. 

The existence of gas accretion to fuel star formation resembles in spirit early prescriptions for 
gas accretion from a hot halo in analytic calculations \citep[e.g.][]{bin77,ree77,sil77,whi78}. 
However, the origin, the morphology, and kinematics of the cold stream model are distinct, 
making this accretion mode the core element of a new paradigm for galaxy formation.

Despite a general acceptance of the cold accretion paradigm from a theoretical perspective, 
this model has been difficult to test empirically. A large body of literature explored the 
possibility of detecting Lyman-$\alpha$ emission from the accreting gas
\citep[e.g.][]{hai00,far01,fur05,dij09,goe10,fau10,ros12}, powered by the potential 
energy of gravitational infall. Unfortunately, predictions for the surface brightness are 
exponentially sensitive to the conditions of the gas (e.g. temperature) and the signal may 
be confused by other sources of Lyman-$\alpha$ photons (e.g. ionization by stars or AGNs, 
that is active galactic nuclei, and scattered radiation). Furthermore, accurate modeling 
requires the solution of coupled hydrodynamic and radiative transfer equations 
\citep[see e.g.][]{ros12}, which is at 
present computationally expensive. To date, no compelling detection of the streams in 
emission has been achieved, although some tantalizing \lya\ observations of filamentary 
structures around high-redshift galaxies have been reported \citep[e.g.][]{can12,rau13,hen13}.

An alternate approach toward direct detection is to observe the cool gas via \ion{H}{1} 
absorption arising from gas that is confined inside or in proximity to dark matter halos, 
within the so-called circumgalactic medium (CGM). High-resolution hydrodynamic simulations 
of galaxy formation predict that cold streams should be manifest as strong absorption systems 
with column densities $\mnhi \ge 10^{17.2} \cm{-2}$, such that they are optically thick 
to Lyman continuum radiation \citep[e.g.][]{fau11,fum11,van12,goe12,she13}. 
Blind surveys along quasar sightlines for these so-called Lyman limit systems (LLSs) 
thus provide, in principle, a test for this scenario. 

\input{table_prop.tex}

One approach is to compare the incidence of optically
thick gas \citep[e.g.][]{pow10,ome13,fum13}, against global estimates for
cold streams in the population of $z>2$ galaxies that are predicted to
contain them \citep{alt11,rah13a}. For instance, simulated massive galaxies with virial
masses $M_{\rm vir}\gtrsim 10^{11}~\rm M_\odot$ at $z\sim3$ do not
account for the entire population of LLSs alone, but consistency
between models and observations could be achieved with an
extrapolation to lower masses \citep{fum11,van12,fum13}. However, a detailed
comparison to theoretical predictions is limited by the fact that these blind 
surveys, by construction, do not directly relate these absorbers to the galaxies 
and dark matter halos that they arise from.

The much more direct approach is to search for signatures of cold
accretion in the vicinity of the $z\sim 2-3$ galaxies that are expected
to host them. Analysis of the stacked spectra constructed from
galaxies lying background to $z\sim 2.5$ star-forming galaxies (the
Lyman break galaxies or LBGs) provide one such test \citep{ste10}, and
models of star-forming galaxies being fed by cold streams appear to
match the average \ion{H}{1} absorption to impact parameters of at
least $\sim 100$\,kpc \citep{fum11,she13}. However, such stacking
analyses can only measure the average equivalent width of \ion{H}{1}
absorption and the flatness of the curve of growth unfortunately dictates
that this method is mostly sensitive to kinematics and only weakly 
dependent on the total amount of absorbing material. 

Ideally, one should probe LBGs with individual sightlines, at 
sufficiently high signal-to-noise ratio and resolution to characterize 
the column densities of absorbers that give rise to cold streams. 
\citet{rud12} have reported on the \nhi\ values measured in
10 quasar sightlines passing within 100\,kpc of a foreground LBG.
They found evidence for optically-thick gas from the CGM in 3 cases.  The implied
covering fraction (here defined as the area subtended by optically-thick gas 
divided by a reference area) is $f_{\rm c} = (30 \pm 14)\%$ within the virial radius (\rvir).
And while future efforts will undoubtedly increase the samples of LBGs 
\citep[e.g.][]{cri11}, building up the data sets of $\sim 100$ sightlines 
required to make robust statistical measurements within \rvir\ will be 
extremely telescope time intensive. 

Recently, \cite{pro13} have expanded on previous efforts \citep{hen06b,hen07,pro09} to 
measure the incidence of optically-thick gas in the CGM of massive galaxies, 
specifically those hosting $z\sim 2$ quasars.  
Using pairs of quasars, they probed the halo gas that is physically associated to a
foreground quasar host galaxy using a background sightline.  Remarkably, this experiment 
reveals a high $f_{\rm c}$, in excess of $60\%$, for sightlines passing within the 
estimated virial radii of these massive galaxies ($\sim 150$\,kpc).  Furthermore, the gas is 
enriched in heavy elements, showing large equivalent widths of low-ion absorption 
(e.g.\ \ion{C}{2}~1334).

The strong clustering of $z\sim 2$ quasars indicates that they are hosted by massive dark 
matter halos $M_{\rm vir} \sim 10^{12.5}\msol$ \citep[e.g.]{whi12}, more than three times 
larger than the typical dark halos hosting LBGs with $M_{\rm vir}\sim 10^{12}~\rm M_\odot$ 
\citep{ade05} at $z\sim 2$. In the current picture of cold accretion, it is believed 
that at these high masses, virial shocks become stable and the CGM of such halos will 
become increasingly dominated by gas that is heated to about the virial temperature 
\citep{dek06}. Qualitatively, one would therefore expect that more massive dark matter 
halos have lower covering fraction of cold gas. For this reason, the results of 
\citet{pro13} are very surprising, being indeed opposite to this naive expectation. 

Motivated by this development, we expand our previous study of 
absorption line systems in the CGM of simulated galaxies \citep{fum11} by 
focusing on the properties of optically-thick gas in a larger suite of 
AMR simulations that have been presented in \citet{cev10}, \citet{cev12}, and 
\citet{dek13}. This new library increases by a factor of three the sample presented in 
\citet{fum11} and includes for the first time galaxies hosted within massive dark matter 
halos ($M_{\rm vir} > 10^{12} \msun$) at $z\sim 2$.
Following our previous work, we include in these simulations recipes for 
star-formation and its feedback and we post-process the outputs with radiative 
transfer calculations to estimate the ionization state of the hydrogen in the 
halos (Section \ref{sims}).

The aim of this paper is to characterize from the theoretical point of view the distribution 
of the neutral hydrogen in cold-stream fed galaxies over more than a decade of halo mass 
(Section \ref{mass}) and to perform direct comparisons to the new observational 
results derived in quasar-galaxy and quasar-quasar pairs, focusing on the incidence 
of optically-thick gas surrounding massive galaxies at $z \sim 2$ (Section \ref{qso}). 
Further, since we will show that the current sample of quasar-galaxy pairs is too small for 
conclusive comparisons with simulations, we propose an additional direct 
test of the cold-stream paradigm by introducing the formalism to compute the 
auto-correlation function of LLSs, a quantity that can be used for statistical 
investigation  of the spatial distribution of optically-thick gas in the CGM 
(Section \ref{future}). The summary and conclusions follow in 
Section \ref{concl}. Throughout this work, for consistency with the parameters 
used in the numerical simulations, we adopt a standard $\Lambda$CDM 
cosmology as described by $\Omega_{\rm m} = 0.27$, $\Omega_\Lambda = 0.73$, 
$\Omega_{\rm m} =0.045$, $h=0.7$ and $\sigma_8=0.82$ \citep{kom09}.

\begin{figure*}
  \centering
  \includegraphics[scale=0.9]{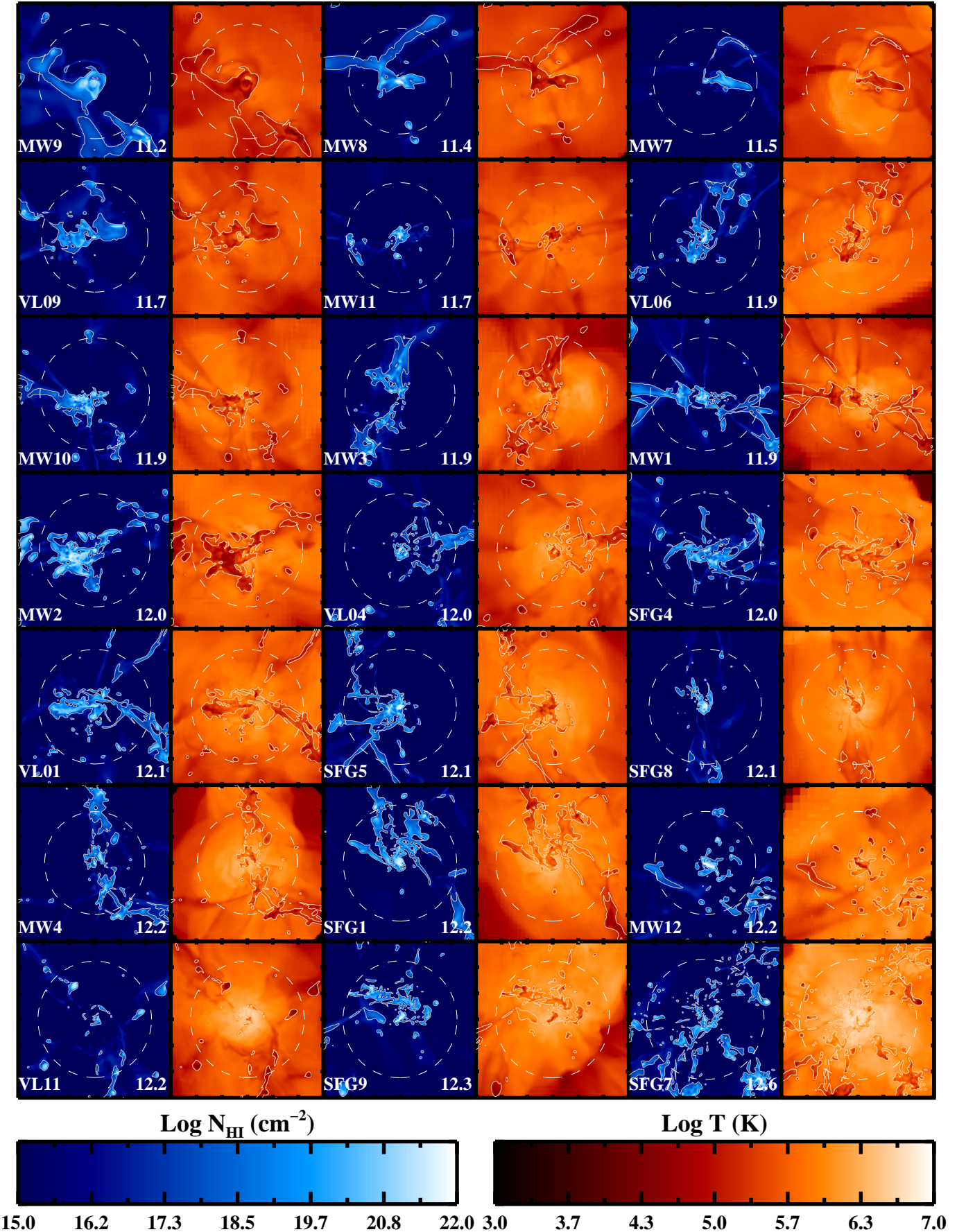}
  \caption{Gallery of the neutral hydrogen and temperature properties of the
    21 halos at $z\sim 2$ that are included in this analysis. For each galaxy we 
    display two panels: on the left, we show the projected neutral hydrogen column density 
    maps with a blue color scheme, while on the right we show the corresponding temperature 
    maps weighted by hydrogen column density with a red color scheme. 
    Regions with $N_{\rm HI} \ge 10^{17.2}~\rm cm^{-2}$ are enclosed in contours. 
    The virial mass is increasing from the top left to the bottom right, as labeled 
    in the bottom right corner of each panel. The virial radius is instead 
    marked by a white circle. Each panel is $\sim 3$\rvir\ on a side.}\label{fig:allgal}
\end{figure*}

\section{Simulations and radiative transfer post-processing}\label{sims}

We present the analysis of 21 galaxy halos at redshift $z \sim 3$ and 
$z \sim 2$, the properties of which are summarized in Table \ref{tab:galprop}.
In this section, we only briefly summarize the numerical techniques used to 
produce the final models. Additional information on these simulations can be found 
in \citet{cev10}, \citet{cev12}, and \citet{dek13}. The procedures adopted for the 
radiative transfer post-processing have been presented in \citet{fum11} and are 
further discussed in Appendix \ref{ratraapp}.

\subsection{Hydrodynamic simulations}

Each halo has been selected from a larger cosmological box and re-simulated with the 
adaptive mesh refinement (AMR) hydro-gravitational code ART \citep{kra97,kra03}. 
The dark matter particle mass is $5.5 \times 10^5 \msun$ and the cell size on the 
finest level of refinement ranges between $35-70$ pc.  At this resolution, densities of 
$n_{\rm H}\sim 10^3~\rm cm^{-3}$ can be reached. In these simulations, refinement occurs when 
the mass in stars and dark matter inside a cell is higher than $2\times10^6~\rm M_\odot$
(i.e. three times the dark matter particle mass), or the gas mass is higher than 
$1.5\times 10^6~\rm M_\odot$.
The ART code incorporates the principal physical processes that are relevant for galaxy 
formation, including gas cooling and photoionization heating, star formation, metal enrichment 
and stellar feedback \citep{cev09,cev10}. Both photo-heating and radiative cooling are 
modeled as a function of the gas density, temperature, metallicity, and UV background (UVB). 
During this calculation, self-shielding of gas is crudely modeled by suppressing the UVB 
intensity to $5.9 \times 10^{26}~\rm erg~s^{-1}~cm^{-2}~Hz^{-1}$
above hydrogen densities $n_{\rm H}=0.1$ cm$^{-3}$. Stochastic star formation occurs at a rate 
that is consistent with the Kennicutt-Schmidt law \citep{ken98} in cells with gas temperature 
$T \le 10^4~\rm K$ and densities $n_{\rm H} \ge 1~\rm cm^{-3}$, but more than half of the 
stars form at $T \lesssim 300~\rm K$ and $n_{\rm H} \ge 1~\rm cm^{-3}$. 

To model feedback processes related to star formation, both the energy from stellar 
winds and supernova type II explosions are injected in the gas at a constant heating rate 
over 40 Myr, while the energy injection from supernovae type Ia is modeled 
with an exponentially declining heating rate with a maximum at 1 Gyr. Cooling is never 
prevented in these simulations and powerful outflows originate in regions where the thermal 
heating due to supernovae and stellar winds overcomes radiative cooling.
In some cases, galactic outflows in these simulations reach high velocities, from few 
hundreds $\rm km~s^{-1}$ to a thousand $\rm km~s^{-1}$
\citep[][]{cev09}, but the mass loading factor is on average low
($\eta \sim 0.3$ at 0.5\rvir). Star formation also enriches the interstellar 
medium (ISM) following the yields of \citet{woo95} and the \citet{mil79} initial mass 
function (IMF).

These simulations are able to reproduce the basic scaling relations observed in high 
redshift galaxies \citep[see][]{cev10}. Nevertheless, because of the limited ability of the 
adopted subgrid prescriptions to model the complex baryonic processes that are associated 
to star formation and feedback, these simulations produce a factor of $\sim 2$ higher 
stellar mass and lower gas fractions by $z\sim 2$ \citep[see a detailed discussion in][]{dek13}.
Further, these simulations do not model feedback from an AGN. However, 
it has been shown by \citet{van12} and \citet{she13} that most of the optically-thick
gas resides in filamentary structures associated to cold gas infall rather than 
in outflowing gas.
Clearly, comparisons with other simulations that include different 
recipes for star formation and stellar winds are needed to verify the extent to which 
the results presented in this paper can be generalized, although at present there are only 
very few zoom-in simulations of halos with $M_{\rm vir} \ge 10^{12}\msun$ in the literature.
Given the high resolution and the fact that AMR codes should capture most of the 
large-scale hydrodynamic processes that are relevant for galaxy formation,
these simulations are among the best models currently available to investigate the 
distribution of hydrogen that originates from the cold streams that feed 
galaxies at high redshifts. We refrain instead from the analysis of the metal 
distribution and gas kinematics, two quantities that are most likely sensitive to 
the adopted feedback prescriptions \citep[e.g.][]{she13}.

\subsection{Hydrogen neutral fraction}

The ionization state of the gas in these simulations is computed in post-processing, 
under the simplistic assumption that the relevant time scales in the radiative transfer 
problem are shorter than the relevant time scales that govern the hydrodynamic equations. 
This approach however neglects the effects of radiative transfer on the hydrodynamics.
Changes in the temperature and ionization fraction of the gas could in fact alter, for 
instance, the properties of the cooling function and the gas pressure 
\citep[cf.][]{fau10,ros12}.

For each AMR cell, we compute the neutral fraction $x_{\rm HI}$ for atomic hydrogen
with a Monte Carlo radiative transfer code, as detailed in Appendix \ref{ratraapp}. 
Both ionization due to electron collisions
and photons are included at equilibrium, but we neglect 
the ionization of helium. Because local sources of radiation are important contributors to 
the ionization of optically-thick hydrogen \citep{fum11,rah13}, 
in addition to the extragalactic UVB from \citet{haa12}, 
we include in our models the radiation from local stellar particles following a Kroupa 
IMF \citep{kro01} and we account for the presence of dust as described in \citet{fum11}. 
Our radiative transfer technique has been validated 
through one of the tests presented in \citet[][see Appendix \ref{ratraapp}]{ili06}. 
Further, the escape fraction from the galaxy disks at the virial radius in these 
simulations is found to be below 10\% \citep{fum11}, consistent with current 
estimates \citep[e.g.][]{nes13}. Finally, independent calculations by 
\citet{rah13} have shown consistency with the results presented in our previous work
\citep{fum11}.

\begin{figure}
  \centering
  \includegraphics[scale=0.38]{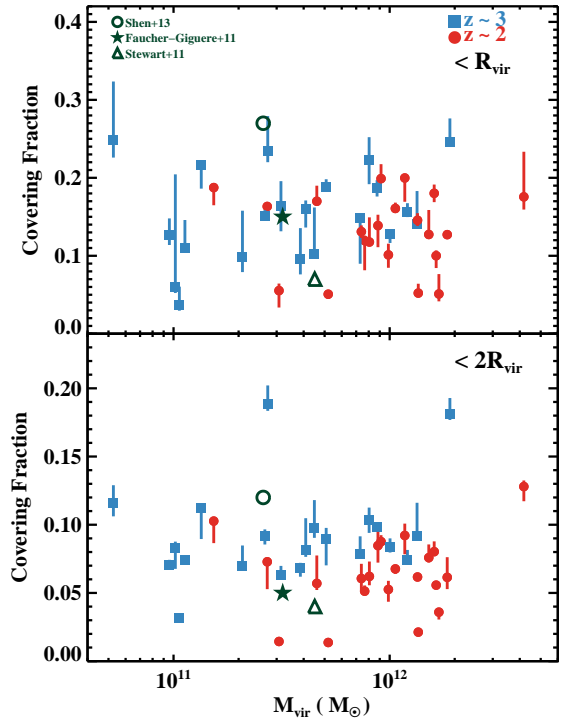}
  \caption{The covering fraction of optically-thick neutral hydrogen as measured 
    within the virial radius (top) and twice the virial radius (bottom). For each simulation, 
    the data points and the error bars represent values measured along three orthogonal 
    directions. Simulations at $z\sim 3$ are shown with blue squares, while models 
    at $z\sim 2$ 
    are shown with red circles. In each panel, we also display the covering fractions 
    from simulations in the literature (green symbols). Open symbols 
    are used for models without detailed radiative transfer post-processing. 
    Simulated galaxies exhibit a wide range of covering fractions, mildly decreasing 
    at fix halo mass from $z\sim3$ to $z\sim2$.}\label{fig:cfall}
\end{figure}

\section{The mass dependence of the neutral hydrogen covering fraction}\label{mass}

In this section, we investigate the mass dependence of the covering fraction of neutral 
hydrogen in these simulations. To this purpose, we generate maps of the neutral hydrogen 
column density in cylinders of radius 2\rvir\ and height 4\rvir. For each 
simulated galaxy, we generate three projections along the three orthogonal axes that 
are naturally defined by the AMR grid. The resolution of the projected maps is comparable to 
the resolution of the smallest cell in each simulation.  For visualization purposes, we also
generate temperature maps, which we construct similarly to the \nhi\ maps 
by averaging the temperature of each cell along the line of sight with weights 
that are proportional to the total column density of hydrogen.
Figure \ref{fig:allgal} presents a gallery of these maps for the $z\sim 2$ galaxies.

\subsection{Cold gas and the critical halo mass}

Simply by inspecting Figure \ref{fig:allgal}, one can already infer the basic CGM properties 
of simulated $z\sim 2$ halos. Across one decade in virial mass 
($M_{\rm vir} \sim 2\times 10^{11} - 4 \times 10^{12}~\rm M_\odot$), 
the average temperature of the lower column density gas 
($N_{\rm HI} \lesssim 10^{16}~\rm cm^{-2}$) is increasing from a few $10^5~\rm K$ to a 
few $10^6~\rm K$. However, at all masses, pockets and narrow filaments
of cooler ($T\lesssim 10^5~\rm K$) and higher column density 
($N_{\rm HI} \gtrsim 10^{17}~\rm cm^{-2}$) gas persist within and beyond the virial radius.

More quantitatively, the volume averaged temperature within the virial radius at $z\sim 2$
is found to increase from $\langle T \rangle \sim 4 \times 10^5~\rm K$ at
$M_{\rm vir} \sim 3 \times 10^{11}~\rm M_\odot$ to 
$\langle T \rangle \sim 3 \times 10^6~\rm K$ at $M_{\rm vir} \sim 4 \times 10^{12}~\rm M_\odot$.
We exclude galactic gas in this calculation by ignoring regions inside 0.15\rvir. 
For halos with virial masses 
$M_{\rm vir}\sim 5\times 10^{11} - 4 \times 10^{12}~\rm M_\odot$,
which bracket the critical halo mass for the formation of stable virial shocks, 
$\langle T \rangle$ is consistent with the predicted post-shock temperature 
$T \sim \frac{3}{8} T_{\rm vir}$, where $T_{\rm vir}$ is the virial 
temperature \citep{bir03,dek06}. 
Virial shocks are also visible in some of the temperature maps presented in 
Figure \ref{fig:allgal}. A similar trend is found in simulations at $z\sim 3$, with  
$\langle T \rangle (z=3) \sim 1.3~ \langle T \rangle (z=2)$
at fixed halo mass, as expected from the redshift dependence of the virial scaling relations.

As already noted in Figure \ref{fig:allgal}, despite the increasing $\langle T \rangle$ 
as a function of halo mass, filaments of cooler and denser material are evident 
in the CGM of even the most massive halos. For gas to exhibit an appreciable fraction 
of neutral hydrogen in absorption, 
typical temperatures have to be $T\lesssim 3\times 10^4~\rm K$, while the volume density 
needs to be $n_{\rm H}\gtrsim 0.003~\rm cm^{-3}$ \citep[e.g.][]{fum11}. Since gas slabs with these 
properties become optically-thick to the incident Lyman continuum radiation, LLSs 
that are relatively straightforward to identify in quasar spectra conveniently 
trace hydrogen with these physical conditions. Therefore, we restrict our analysis of 
the cool halo gas to column densities of $N_{\rm HI}\ge 10^{17.2}~\rm cm^{-2}$, which we can 
also compare to existing observations.

\begin{figure}
  \centering
  \includegraphics[scale=0.38]{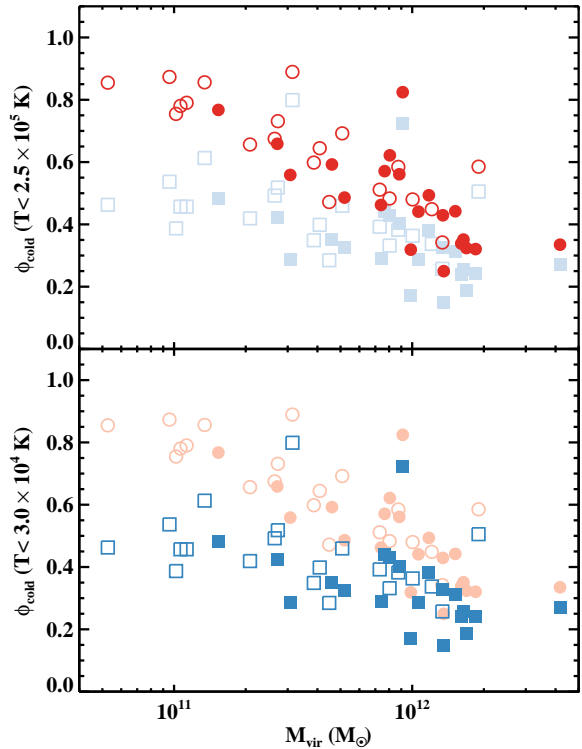}
  \caption{The mass fraction of gas at the {\it instantaneous} temperature of
    $T<2.5\times 10^5~\rm K$ (top; circles) and $T<3\times 10^4~\rm K$ (bottom; squares) 
    that is enclosed within $0.15 R_{\rm vir} < r < R_{\rm vir}$. 
    Simulations at $z\sim3$ and $z\sim2$ are shown with 
    open and filled circles, respectively. The distribution from the top panel is 
    overlaid to the distribution in the bottom panel (and vice-versa) to facilitate 
    comparisons. 
    The fraction of gas with $T<2.5\times 10^5~\rm K$ 
    decreases with increasing halo mass, while the mass fraction of the colder 
    gas ($T<3\times 10^4~\rm K$) at any given redshift is only weakly dependent 
    on mass.}\label{fig:fcold}
\end{figure}

Figure \ref{fig:cfall} summarizes the covering fractions of
optically-thick gas in the CGM of the 21 simulations under
examination, both at $z\sim 2$ and $z\sim 3$.  
In this paper, we focus on an empirical definition for 
the covering fraction because we aim to extract 
from simulations an observable quantity that can be directly compared to observations.
Our covering fraction encompasses all gas that is optically-thick to a background 
source in projection, regardless to its kinematic state \citep[cf.][]{van12}, 
including gas that is associated to the central galaxies.
In fact, in observations, one cannot trivially disentangle 
the contribution of halo gas from the contribution of the outskirts of
galaxy disks. A subtlety arises, however, from the fact that galaxy-quasar pairs or
quasar-quasar pairs are intrinsically rare at very small projected separations and,
whenever possible, in the following we compare observations and simulations using the
observed distribution of impact parameters. Furthermore, since observations cannot 
separate halo gas from gas associated to satellites, we include gas 
within satellite galaxies 
in our definition of $f_{\rm c}$ \citep[see figure 7 in][for estimates of 
$f_{\rm c}$ with and without the contribution of satellites]{fum11}. 
We emphasize that, since $f_{\rm c}$ includes also gas that 
is not infalling, this is an upper limit to the theoretical covering fraction of 
accreting gas within the CGM. Furthermore, given our (arbitrary) definition 
for gas inside galaxies 
($R<r$\rvir\ with $r=0.15$), one can trivially derive a lower limit to the covering fraction 
of halo gas without the galaxy contribution: $f'_{\rm c} \ge (f_{\rm c}/r^2-1)/(r^{-2}-1)$. 
As expected, the correction $f'_{\rm c}/f_{\rm c}$ is large for the few galaxies with small 
$f_{\rm c}$ (like SFG8 or MW11), but minor ($<20\%$) for most of the galaxies with 
$f_{\rm c} \ge 0.1$.

Figure \ref{fig:cfall} shows that the range of $f_{\rm c}$ within the virial radius 
is between $f_{\rm c}\sim 0.05 - 0.2$ at $z\sim 2$. Variations resulting from projection 
effects, albeit quite large in some galaxies, are typically smaller than this scatter,
which reflects instead an intrinsic variation in the gas accretion and
merger history of halos.  This large scatter should discourage one
from generalizing results obtained from a single simulation, which has
often been done in the literature. 
Due to the geometry of the filaments that extend radially outward, the
covering fraction at 2\rvir\ drops between $f_{\rm c} \sim 0.01 -
0.13$, implying that an approximately equal area is subtended by
optically-thick gas within \rvir\ and $R_{\rm vir} < R < 2R_{\rm
  vir}$. Comparing the redshift evolution of individual galaxies, we find only
a modest decrease in the covering fraction that at $z\sim2$ drops to
$\sim 70\%$ of the value measured at $z\sim 3$ within 2\rvir\ ($\sim 80\%$
at \rvir).

Figure \ref{fig:cfall} also shows a lack of any appreciable mass dependence 
of the covering fraction over one decade in virial mass, despite the 
fact that our sample brackets the critical mass of $\sim 5\times 10^{11}~\rm M_{\rm vir}$ 
above which virial shocks become stable \citep{dek06,ocv08}.
A general prediction of cosmological hydrodynamic simulations is that the fraction of 
cold gas decreases as a function of virial mass \citep[e.g.][]{ocv08,ker09b,van11,fau11b,nel13}. 
This fact would naively suggest a lower covering fraction of neutral hydrogen 
in more massive halos, but Figure \ref{fig:cfall} 
illustrate that is indeed not the case for our simulations. 

Gas has been defined ``cold'' differently by various authors, and the accretion 
rates or the ultimate fate of the cold material falling onto galaxies
are extensively discussed -- and highly debated -- in the literature.
The goal of our analysis is not to determine the detailed evolution of
cold gas in galaxies at $z\sim 2-3$, but instead we focus on predicting 
the covering fraction of optically-thick gas around galactic halos at any given time, 
and on understanding its relationship to the mass fraction of cold gas $\phi_{\rm cold}$.  
Predictions for $f_{\rm c}$ are of obvious interest for understanding the origin of LLSs, 
and furthermore this covering fraction is an observable quantity for which  recent measurements 
exist. Thus, here we define cold gas using the instantaneous temperature
at a given redshift, i.e. without considering the past or future thermal 
history of this gas.

To gain further insight into the weak mass dependence of the covering
fraction in our simulations (Figure \ref{fig:cfall}), we compute the
fraction of cold gas $\phi_{\rm cold}$ within the virial radius for
our simulated galaxies at $z\sim 2$ and $z\sim3$, which is shown in
Figure \ref{fig:fcold}. Here, $\phi_{\rm cold}$ is defined as the
ratio of the cold gas mass to the total gas mass within a given
radius. In agreement with previous work, gas within 0.15\rvir\ has been excluded 
from the analysis and from the values listed in Table \ref{tab:galprop} 
to avoid material residing in the galaxy disk. If we define gas as
``cold'' when the temperature is less than $2.5\times 10^5~\rm K$, we
find a trend of decreasing $\phi_{\rm cold}$ with increasing virial mass (top
panel of Figure \ref{fig:fcold}), in qualitative agreement with
previous simulations \citep[e.g.][]{ocv08,ker09b,van11,nel13}. 
However, when we refine the definition of cold gas to include only
hydrogen that is likely to remain neutral when self-shielded from
ionizing radiation (i.e. $T\sim 3 \times 10^{4}~\rm K$; Table \ref{tab:galprop}), 
we find a shallower dependence of $\phi_{\rm cold}$ on halo mass (bottom panel
of Figure \ref{fig:fcold}). 

Thus, in our simulations we observe both an increase in the ``hot'' gas fraction with 
virial mass and a mass-independent $f_{\rm c}$ for optically-thick gas. This result is
in apparent contradiction with the naive expectations based on
previous work which, however, did not directly characterize the mass
dependence of the covering fraction of optically-thick gas at any given redshift, which is
the observable quantity. In other words, the onset of stable virial shocks affects the 
temperature and the mass fraction of gas at $\gtrsim 10^5~\rm K$, without preventing the 
existence of colder and neutral gas pockets in galaxy halos, even for masses above the 
critical halo mass for shock formation. Qualitatively, this is consistent with the idea 
that filaments of cold gas survive above the transition mass at $z \ge 2$ 
\citep{ker05,dek06,ocv08,ker09b}.

Finally, a mass-independent covering fraction
may appear in conflict with recent reports by \citet{ste11a} of a decreasing $f_{\rm c}$
once a galaxy crosses the critical mass for the formation of hot halos. 
However, it should be noted that these authors follow 
the redshift evolution of two halos, finding a drop in the covering fraction 
only for $z<1.5$. Therefore, in light of the previous discussion, we 
interpret the sudden decrease in $f_{\rm c}$ reported by \citet{ste11a} 
as not being simply due to the halo growing beyond the critical mass and the
concomitant presence of shock heated gas. But rather other factors,
including redshift evolution, have to play a role in shaping the
covering fraction seen in these simulations.  Furthermore, it should
be noted that the critical mass does not coincide with exactly the
same halo mass for all galaxies, but instead depends on when the virial
shock is triggered. Values of critical mass can spread over more than
a decade in mass \citep[e.g.][]{ker05}.

\subsection{Comparisons with other simulations}

In Figure \ref{fig:cfall}, we compare the covering fractions measured in our 
simulated galaxies to values from other simulations 
published in the literature. The covering fraction of the 
{\it Eris} halo, simulated at $z\sim 2.8$ by \citet{she13} with an SPH code, is consistent 
with the upper limit of our distribution at $z\sim 3$, although their analysis relies on 
simple approximations for the ionization state of the gas. Similar consistency is found
for the SPH simulation of the Milky Way progenitor B1 by \citet{fau11} at $z\sim 2$
and for the SPH models with virial masses between 
$\sim 3\times 10^{11} -6 \times 10^{11}~\rm M_\odot$ at $z\sim 2$ by \citet{ste11}. 

There seems to be agreement in the covering fractions of halos 
simulated with different numerical techniques (AMR and SPH), 
but this comparison is at the moment rather crude, since it is based on a 
very basic metric.  For instance, the covering fraction may not 
properly reflect the difficulties of classical SPH formulations in 
capturing contact discontinuities and instabilities
\citep[e.g.][]{age07,sij12} or sub-sonic turbulence dissipation \citep{bau12}
that can affect both the properties of hot halos and of cold filaments inside 
massive halos. We now await comparisons with simulations performed with new 
SPH implementations that mitigate these problems \citep{rea12,hop13}.
Moreover, as noted, some of these simulations do not incorporate
a detailed radiative transfer post-processing which is crucial to 
correctly describe the neutral fraction at the column densities relevant to LLSs,
nor do they implement the same prescriptions for sub-grid physics. 
Finally, as previously highlighted, the large scatter in 
$f_{\rm c}$ within our ensemble of simulated galaxies hampers 
a precise comparison simulations of individual halos. 
Future analysis, e.g. from the ongoing AGORA code comparison project, 
will provide a better characterization of the level of agreement between 
various simulations. 

\citet{bir13} compared halos from a cosmological box simulated
with an SPH code and with the new moving-mesh code Arepo, 
without radiative transfer post-processing and at lower resolution. 
These authors concluded that SPH codes 
produce an excess of optically-thick gas around halos of $M_{\rm vir} > 10^{11}~\rm h^{-1}~M_\odot$ 
compared to Arepo simulations. Thus, one may conclude that galaxies simulated with Arepo 
have lower covering fractions than what is found in SPH simulations. 
Distressingly, this would worsen the current tension between numerical calculations and 
observations (Section \ref{qso}). However, before drawing similar conclusions, we prefer to await 
additional comparisons between Arepo and SPH or AMR codes at the high resolutions that 
are comparable to the ones achieved by the simulations presented in Figure \ref{fig:cfall},
once radiative transfer post-processing has been included. 

Finally, we acknowledge that other simulated halos with comparable 
redshifts and masses to those included in this study have been presented 
in the literature \citep[e.g.][]{ros12,hum13}, but because these authors do not 
provide direct information on the covering fraction of optically-thick gas,
these simulations do not appear in Figure \ref{fig:cfall}. Nevertheless, 
the column density maps presented by \citet{ros12} appear in qualitative agreement 
with the maps shown in Figure \ref{fig:allgal}. Also, \citet{hum13} comment on the agreement
between their model and the simulations of \citet{fau11}.

\begin{figure}
  \centering
  \includegraphics[scale=0.32,angle=90]{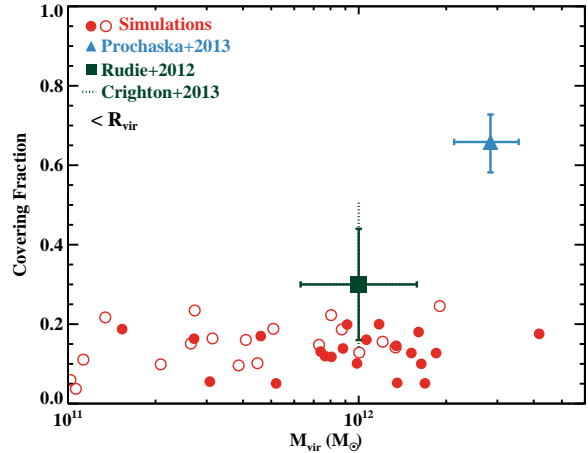}
  \caption{Comparison of the covering fraction of optically-thick gas in simulated and observed 
    galaxies. Simulations at $z\sim3$ and $z\sim2$ are represented by empty and filled 
    red circles, respectively. The observed covering fraction around 
    $z\sim2$ quasar host galaxies is shown with a blue upward triangle, while observations for
    Lyman break galaxies at $z\sim 2.5$ are summarized by a green square. 
    The horizontal error bars reflect the large uncertainty in the inferred halo mass 
    for these objects. Simulations appear to systematically underpredict the observed 
    covering fractions at the highest masses, while current samples are too small for conclusive
    comparisons with LBGs.}\label{fig:cfobs}
\end{figure}

\section{Simulations versus observations}\label{qso}

Having characterized the covering fractions in galaxies at $z \sim
2-3$ from a theoretical point of view, in this section we directly
compare the predictions from our simulations with observations.

\subsection{Covering fractions within \rvir}

In Figure \ref{fig:cfobs}, we show again the simulated covering
fractions within \rvir\ both at $z\sim 2$ and $z\sim 3$, but we now
superimpose measurements of the covering fractions of optically-thick
gas for LBGs \citep[][N. Crighton et al., in prep.]{rud12} and in
quasar host galaxies \citep{pro13}. 

\citet{rud12} have measured the covering fraction of optically-thick gas in a 
sample of 10 LBGs within 100 kpc from a bright background quasar. 
Typical halo masses for LBGs are inferred by comparing the observed 
clustering of galaxies to the clustering of dark matter halos in 
numerical simulations. In the following, we assume the mass interval 
of $10^{11.8} < M/{\rm M_\odot}< 10^{12.2}$ at $z\sim 2$ from 
\citet{ade05} \citep[see also][]{con08}, where the uncertainty in the 
halo mass reflects the errors on the measured correlation function.
However, different determinations may suffer from larger systematic 
uncertainties \citep[see e.g.][]{bie13}. 
Assuming $R_{\rm vir}\sim 90~\rm kpc$ for galaxies at this mass, \citet{rud12}
find $f_{\rm c} = 0.30 \pm 0.14$ within 68\% confidence interval inside the 
virial radius. A similar analysis by N. Crighton et al. (in prep.) yields a 
comparable covering fraction, with slightly larger error bars. 

A subset of our simulated galaxies or the {\it Eris} simulation by \citet{she13} 
approach the observed value. However, we emphasize that 
this comparison is subject to the uncertainties of the subgrid physics  
included in these simulations (see Section \ref{secwind}).
As a population, the covering fraction in 
simulations ($f_{\rm c} = 0.15 \pm 0.06$) is a factor of 2 
lower than what suggested by observations \citep[cf.][]{rud12}, 
but is nevertheless consistent given the large 
error bars. The mean covering fraction in simulations is in fact 
in formal agreement with observations, lying within the 68\% confidence interval.
This comparison therefore highlights how current samples of LBGs at $z\sim 2-3$ 
in proximity to background quasars are too small to conclusively establish whether 
there is inconsistency between simulations and observations, limiting 
our ability to robustly test current theories for gas accretion onto galaxies.

The situation is instead different at larger masses, as shown in
Figure \ref{fig:cfobs}.  Using a sample of 74 quasar pairs,
\citet{pro13} have measured the covering fractions of optically-thick
gas in the surroundings of $z\sim 2$ quasar host galaxies at projected
separations ranging from 30 to 300 kpc.  Assuming a typical halo mass
of $(2.85\pm 0.71)\times 10^{12}~M_\odot$ for the quasar host halos
deduced from the clustering measurements of \citep{whi12}, and a
corresponding virial radius of $\sim 150~\rm kpc$, 27 optically-thick
systems are found along the 41 sightlines that sample the halos within
\rvir.  The inferred covering fraction is therefore $f_{\rm c} = 0.67
\pm 0.07$ (68\% confidence interval). As evident from Figure
\ref{fig:cfobs}, this covering fraction significantly exceeds the
values measured in these simulations.

\begin{figure}
  \centering
  \includegraphics[scale=0.32,angle=90]{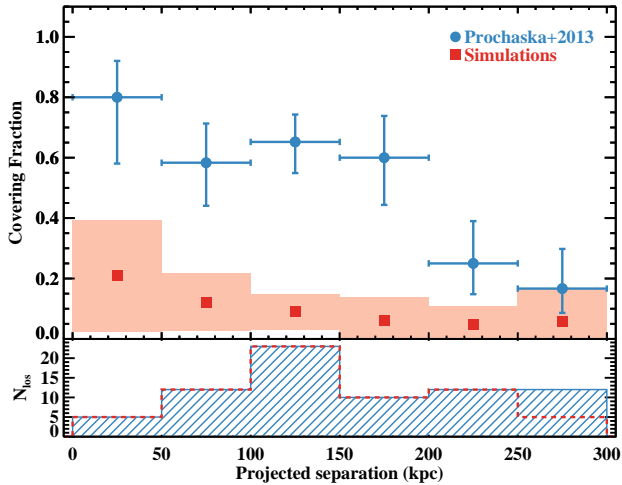}
  \caption{Radial covering fractions for the five most massive halos simulated at 
    $z \sim 2$ (red squares) 
    and in observations of quasar host galaxies (blue circles). 
    Error bars in the covering fractions around quasars represent the 68\% confidence Wilson 
    score interval. For simulations, instead, we show the mean covering fractions 
    computed in 1000 trials, with the standard deviation shown by the red shaded area.
    The bottom panel shows the number of sightlines in each radial bin in simulations 
    (red dashed histogram) and observations (shaded histogram). Due to the finite box size 
    of simulations, the outermost radial bin is undersampled compared to observations. Within 
    $\sim 200~\rm kpc$, simulations show a significantly lower covering fractions than what is 
    found in quasar host galaxies.}\label{fig:qsocf}
\end{figure}

\subsection{The radial dependence of $f_{\rm c}$}

The \citet{pro13} quasar pair sample is sufficiently large to enable
measurements of the covering fraction as a function of projected
separation from the foreground quasar.  A comparison between
observations and simulations for the 5 most massive halos 
above $M_{\rm vir}=1.6\times 10^{12}~\rm M_\odot$ in our
sample (MW12, SFG1, SFG7, SFG9, and VL11) is presented in Figure
\ref{fig:qsocf}. The mean virial mass in this subset is $M_{\rm
  vir}=2.2\times 10^{12}~\rm M_\odot$ (median  
$M_{\rm vir}=1.7\times 10^{12}~\rm M_\odot$), 
comparable to the typical halo mass of quasar host galaxies. In this figure, 
the covering fractions and the corresponding 68\% Wilson confidence intervals 
deduced from the observations are shown in bins of projected separation between the
foreground quasar and the background quasar sightline. We have assumed
the quasar resides at the center of its host dark matter halo.
For a consistent comparison, we generated 1000 realizations
of the same experiment conducted by \citet{pro13}, but using
our simulations. For each trial, we randomly sample the five simulated
halos along three orthogonal axis using 74 sightlines at the exact
same set of impact parameters as the observed quasar pair sample.

The mean of the covering fractions computed for each bin are compared
to the measurements in Figure \ref{fig:qsocf}, where we also show the
standard deviation of the distribution of covering fraction in each bin
measured from our ensemble of realizations. Because of the limited 
size of our simulation cube (4\rvir\ on a side), the bin at the largest
impact parameters is slightly undersampled in our mock observations
(see bottom panel of Figure \ref{fig:qsocf}).  The inconsistency
between observations and simulations is readily apparent for
separations $\lesssim 200~\rm kpc$. Given the limited number of
sightlines within 50 kpc, the large difference between the observed 
and simulated covering fractions in the innermost bin is not 
statistically significant. However, in the interval
between $50-200~\rm kpc$, all the simulated covering fractions
are significantly below the observations, lying
outside the 95\% confidence interval measured for the quasar pair data. This
striking discrepancy is also evident from the fact that there were 
37 optically-thick systems found in the 74 observed sightlines, whereas
we never found 37 or more such absorbers in our 1000 trials sampled at the 
same 74 impact parameters. 

The picture that clearly emerges from this comparison is that the 
basic cosmological processes responsible for the assembly of massive
galaxies, and particularly gas inflows, do not produce a sufficiently high
covering fraction of optically-thick gas to explain the high value
observed around quasar host galaxies. This is especially
true given that our post processing radiative transfer does not 
include the effect of the additional ionizing photons from the quasar itself, 
which would even further reduce the covering fractions deduced from the simulations
\citep{hen06b,hen07,pro13}.

\subsection{Impact of feedback mechanisms on comparisons between simulations and observations}\label{secwind}

The foregoing analysis reveals that our understanding of the gas distribution in 
the massive galaxies that host quasars is incomplete. In this section, we briefly 
speculate on possible causes for the discrepancy highlighted by our study, focusing 
first on feedback mechanisms.

The simulations included in this study (as other simulations discussed in the recent 
literature) are imperfect models of our Universe, particularly because of the weak or 
ad-hoc implementation of feedback. As discussed in Section \ref{sims}, the average 
mass loading factors of the winds in these simulations is low, $\eta\sim 0.3$ at 
0.5\rvir. Therefore, these simulations overestimate the amount of stars formed by 
$z\sim2$ by a factor of  $\sim 2$, and consequently underpredict the gas fractions 
within the galaxy disks. This fact may impact the simulated properties of the CGM 
in several ways. 

For instance, strong outflows would prevent gas to be locked into stars at 
high redshift, and additional material may then available for later accretion 
\citep[cf.][]{opp10}. At the same time, stronger outflows may interact with 
the accreting material shaping its structure \citep[see a discussion in][]{fau11b,pow11}.
Further, a stronger implementation of stellar feedback 
\citep[see e.g.][]{sti12,she13,cev13}, or an additional form of feedback from 
the central AGN, may be the astrophysical process that is needed to boost 
the covering fractions of optically-thick gas in these simulations.
Besides alleviating or even resolving the tension between observations 
and simulations, feedback processes may also be required to reproduce the large
equivalent widths of metal lines (e.g. for \ion{C}{2}) that have been
found within the virial radius of quasar host galaxies \citep{pro13}.

However, detailed absorption line modeling and analysis of the
physical properties of a single quasar absorption system in
\citet{pro09} indicated that the enriched gas detected in the quasar
CGM was unlikely to represent material ejected from the AGN. 
Furthermore, as shown by \citet{van12} and \citet{she13}, the majority 
of the cross section of optically-thick hydrogen lies in cold filaments, 
with only a small contribution originating in cold gas entrained within 
outflows. For these reasons, stronger feedback implementations 
that generate mostly hot winds may not significantly boost the cross section of 
optically-thick gas. Different implementations in which a larger fraction of cold material 
is entrained in the outflowing gas may be required to increase the cross 
section of optically-thick gas. Unfortunately, most of the relevant astrophysical 
and hydrodynamic processes that occur in winds are currently not fully resolved 
by cosmological hydrodynamic simulations \citep[see][]{pow11,jou12,hop12,cre13}.

We also emphasize that resolution may play a significant role in shaping the structure of 
optically-thick gas in simulations, regardless to the adopted feedback models. 
At progressively lower resolution, high density peaks are smoothed out, and thus 
structures of size comparable to the grid cells are not properly capture in these 
simulations. Thus resolution may affect the resulting covering fraction 
directly, e.g. we could be missing small clumps of optically-thick gas, or 
indirectly, e.g. by altering the structure of the medium through which ionizing 
photons propagate during our radiative transfer post-processing.

\subsection{Additional causes for the discrepancies between simulations and observations}

Besides incomplete physics in our simulations, other reasons
can be invoked to account for the current inconsistency between
simulations and observations at the high-mass end. Because of our 
limited simulation volume, the optically-thick gas modeled in these
simulations resides within $2R_{\rm vir}$ from the center
of the halo. Conversely, the \citet{pro13} analysis considered
a velocity interval of $\pm 1500 \rm km~s^{-1}$, which was required
because of significant uncertainties in the quasar redshifts \citep{ric02}. 
This velocity interval corresponds to $\pm 45~\rm Mpc$ along the line-of-sight, 
and thus optically-thick systems detected at small projected distances 
(e.g. within \rvir) from a foreground quasar could in fact lie at larger line-of-sight 
separations, and hence larger physical separations than we have considered. 

Given the observed number of LLS per unit redshift at $z\sim 2$ \citep{ome13}, 
the probability to intercept a random LLS from the cosmic background over
such a small redshift path is however negligible compared to the large
covering factors observed.  However, if quasars reside at the center
of larger scale structures, such as group of galaxies, which are each
surrounded by optically-thick halo gas, then the observed covering
factor may include a contribution from optically-thick absorbers at
distances larger than the $2R_{\rm vir}$ that we have considered. This
effect needs to be investigated with simulations of larger
cosmological volumes. However, we speculate that absorbers distance
larger than $2R_{\rm vir}$ can ease but not resolve the discrepancy
between observations and simulations. Indeed, \citet{pro13} measure a
dropoff of the $f_{\rm c}$ with impact parameter for $r > 200\,{\rm
  kpc}$ (Figure \ref{fig:qsocf}), which suggests that optically-thick gas 
is mostly contained in proximity to the central galaxy and  
argues against a large contribution to the covering
fraction from Mpc scales. 

Finally, if quasars mark a particular phase in the life of a galaxy 
in which the AGN activity is triggered by mergers \citep[e.g.][]{san88,dim05,hop05},
the observations of quasar pairs may provide only a biased view of the halo gas in 
massive galaxies. However, processes other than major mergers may be responsible for feeding 
AGNs, in particular at high redshifts \citep[e.g.][]{dav09,cio10,bou11,cis11,dim12}. 
Furthermore, at the typical bolometric luminosity of the quasar pairs 
($L_{\rm bol} = 10^{45.5}-10^{47}~\rm erg~s^{-1}$), observations implies 
a star formation rate of $\sim 10-100~\rm M_\odot~yr^{-1}$ \citep[e.g.][]{tra10}
which is comparable to the star formation rates observed in matched populations 
of non-active star forming galaxies \citep[e.g.][]{sha10,san12,mul12,har12,ros13}.
Thus, at present, there is no clear indication that quasar pairs 
reside in a population of halos that are systematically different than those
described by our simulations.

\section{A statistical view of the circumgalactic medium}\label{future}

As shown in Section \ref{qso}, samples of LBG-quasar pairs are currently 
too limited in size for conclusive comparisons with simulations. 
In the second part of this paper, we therefore introduce the formalism 
for measuring the auto-correlation function of LLSs (Section \ref{llslls}), 
which is based on an extension of the formalism used to measure the galaxy-LLS 
cross-correlation function (reviewed in Section \ref{gallls}). The advantage of this
experiment is to exploit larger samples of quasar pairs to statistically map the 
distribution of optically-thick hydrogen around galaxies at $z\sim 2-3$,
avoiding the telescope-intensive task of finding many galaxy-quasar pairs. 

\subsection{The galaxy-LLS correlation function}\label{gallls}

In Figure \ref{fig:qsocf}, we have shown the radial dependence of the covering fraction in 
quasar host galaxies. This quantity, which is particularly useful to investigate the spatial 
extent of the CGM around galaxies of a given halo mass, can be recast in terms of the 
galaxy-LLS cross-correlation function $\xi_{gl}(r)$ \citep[see][]{hen07,pro13b}. The 
cross-correlation function contains the same information as the covering fraction, 
but it has the advantage of directly comparing gas around galaxies 
to the cosmic background abundance of optically-thick hydrogen absorbers that are
intercepted randomly as intervening LLSs. Thus, it directly 
quantifies the spatial scales for which a statistically significant excess of
optically-thick absorption is detected around galaxies. 

This cross-correlation function can also be compared to, e.g. the auto-correlation function of 
the galaxies themselves as well as the underlying dark-matter distribution, to help further 
constrain the distribution of CGM gas relative to the large-scale structure 
\citep[e.g.][]{sel00,wei04,coo02}. Indeed, 
cross-correlation functions between galaxies and absorbers have already been studied in 
the literature. For instance, \citet{bou04} and \citet{coo06} measured the correlation 
between LBGs and damped Lyman-$\alpha$ systems, while \citet{hen07}, \citet{fon13}, 
and \citet{pro13b} measured the clustering of either LLSs or the Lyman-$\alpha$ forest 
around quasars. Also, \citet{tin08}, \citet{wil08}, and \citet{ade05b} studied the 
correlation between galaxies or quasars and metal absorption lines.

With the exception of \citet{hen07} and \citet{pro13b}, all previous work has 
measured clustering on scales larger than $\sim 1~\rm Mpc$, and these larger scale 
clustering measurements 
constrain the dark matter halos hosting absorbers \citep{tin08}. However, as we will argue below, 
the small-scale clustering  (i.e. scales comparable to the virial radius) or ``one-halo'' term 
has the potential to provide a very sensitive test for simulations of the CGM around galaxies. 
In what follows, we briefly review the formalism to compute the galaxy-LLS correlation function, 
closely following the discussion in \citet{hen07}. We then show predictions of $\xi_{gl}(r)$ 
computed from numerical models which we will then compared to measurement for the 
LLS auto-correlation function.

\subsubsection{Formalism}

For a given a population of galaxies with redshifts $z_0$ that are probed by background
quasars at projected separations $r_\bot$, we describe the 
distribution of optically-thick gas around halos as an excess probability of finding a 
LLS in comparison to random expectation inside a velocity interval $\pm \Delta v$ 
that is centered at the galaxy systemic redshift.

The probability of finding a LLS at random in the corresponding redshift interval 
$\Delta z_{\rm 0} = 2 \Delta v (1+z_0)/c$ is 
$P(\Delta z_0,r_\bot)=\ell(z_0)\Delta z_0$, where $\ell(z)$ is the number 
of LLSs per unit redshift evaluated at $z_0$. This probability, which is 
independent of the  projected separation, expresses the
covering fraction of absorbers from the cosmic background population of random 
intervening LLSs. At a distance $r_\bot$ from a foreground galaxy (or quasar host galaxy), 
the probability of intercepting optically-thick gas is enhanced by clustering 
around the galaxy according to 
\begin{equation}\label{plx}
F_{\rm c} (\Delta z_0,r_\bot) = \ell (z_0) [1+\chi_{\rm gl, \bot} 
  (\Delta z_{\rm 0}, r_\bot)] \Delta z_0\:.
\end{equation}
Here, $\chi_{\rm gl, \bot} (\Delta z_{\rm 0}, r_\bot)$  is the projected galaxy-LLS cross-correlation 
function, which quantifies the excess probability  above the cosmic mean of detecting LLSs near 
the galaxy in the corresponding redshift interval. As we will show below, this probability 
$F_{\rm c}$ is directly related to the covering fraction $f_{\rm c}$ of optically-thick gas around 
galaxies.

The projected correlation function $\chi_{\rm gl, \bot} (\Delta z_{\rm 0}, r_\bot)$ can be 
related to the real-space galaxy-LLS correlation function $\xi_{\rm gl}(r)$ with 
an average over the volume $V=\sigma_{\rm a,c}r_{||}$. Here,  
\begin{equation}\label{rpar}
r_{||} = \frac{c}{H_0} \int_{\Delta z_0} \frac{dz}{\sqrt{(\Omega_{\rm m}(1+z)^3 
    + \Omega_{\rm \Lambda})}} \approx 
\frac{c \Delta z_0}{H(z)}
\end{equation}
in a flat cosmology, and $\sigma_{\rm a,c}$ is the cross section of the absorbing clouds. 
Under the assumption that $r_\bot >> \sigma_{\rm a,c} ^{1/2}$, 
\begin{equation}\label{intchi}
\chi_{\rm gl,\bot} (\Delta z_{\rm 0}, r_\bot) \approx \frac{1}{r_{||}} \int ^{+r_{||}/2}_{-r_{||}/2} \xi_{\rm gl} (r'_{||}, r_\bot) dr'_{||}\:.
\end{equation}

For an ensemble of galaxy/quasar pairs, the projected galaxy-LLS
correlation function $\chi_{\rm gl,\bot}$ can be evaluated in bins\footnote{For 
alternative method for computing the projected correlation function without binning
data see, e.g., \citet{hen07}.} of $r_\bot$ as
\begin{equation}\label{binnedchi}
\chi_{\rm gl,\bot}(\Delta z_{\rm 0}, r_\bot) = \frac{N_{\rm lls}}{N_{\rm ran}}-1\:. 
\end{equation}
Here, $N_{\rm lls}$ is the number of LLSs detected around the galaxies in bins centered on 
$r_\bot$ and $N_{\rm ran}$ is the number of LLSs expected at random for a given $\ell(z)$.
Given measurements of $\chi_{\rm gl,\bot}(\Delta z_{\rm 0}, r_\bot)$, 
one can determine the functional form 
for $\xi_{\rm gl}(r)$ that best describes the observations using Equation (\ref{intchi}). 

\begin{figure}
  \centering
  \includegraphics[scale=0.32,angle=90]{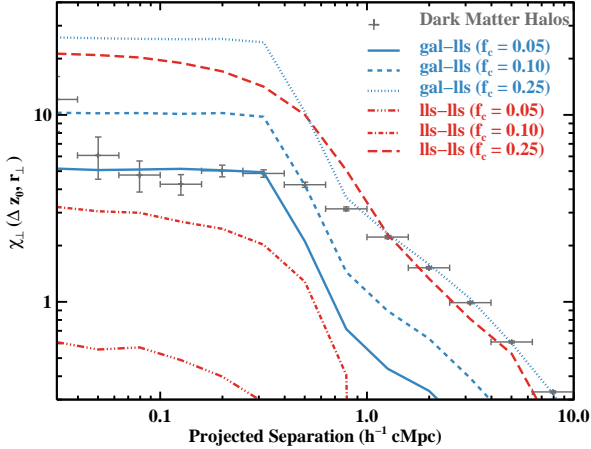}
  \caption{The projected galaxy-LLS cross-correlation functions  
    computed for different covering fractions ($f_{\rm c}=0.25, 0.10, 0.05$) 
    of optically-thick gas (blue lines)
    around dark matter halos with masses 
    $5\times 10^{11} - 5 \times 10^{12}~\rm M_\odot$ within a 250 cMpc/h cosmological box.
    The projected LLS auto-correlation function  
    is shown with a red dashed line for $f_{\rm c}=0.25$, with a red 
    dotted line for  $f_{\rm c}=0.10$, and with 
    a red dashed triple-dotted line for $f_{\rm c}=0.05$. The 
    projected two-point correlation function of dark matter halos is shown 
    by grey crosses. For these calculations, we assume a velocity window 
    $\Delta v = \pm 400~\rm km~s^{-1}$. If LLSs statistically trace galaxies, 
    the LLS auto-correlation function encodes the same information that is 
    contained in the galaxy-LLS correlation function, only smoothed on scales of 
    the gaseous halo.}\label{fig:llslls}
\end{figure}

\subsubsection{Numerical models}

Our goal is to show with simple numerical models how
the LLS auto-correlation function can be used to gain insight into the 
properties of the CGM in comparison to the galaxy-LLS cross-correlation 
function, and not to produce detailed predictions for these two quantities. 
Therefore, we generate simple realizations of a universe in which LLSs are distributed 
around galaxies adopting the following prescriptions. 

The spatial distribution of dark matter halos is given by 
the {\sc rockstar} halo catalogue \citep{beh13} extracted form 
the Bolshoi simulation \citep{kly11}, a dark matter only cosmological 
simulation in a box of 250 cMpc/h (comoving Mpc) on a side.  
We then model the spatial distribution of LLSs by populating dark matter 
halos at $z\sim2$ in the mass interval $5\times 10^{11} - 5 \times 10^{12}~\rm M_\odot$
(consistent with the range explored in the previous sections) with a
varying covering fraction of $f_{\rm c}=0.05, 0.10$, and  $0.25$ within 2\rvir.  
These values can be compared to the results of the hydrodynamic zoom-in
simulations presented in the first part of this paper or to the 
observed values around LBGs \citep[][]{rud12}.

In this way, we obtain realizations of a
universe in which, by construction, a fraction $\ell(z)_{\rm halo}
\propto 4 \pi R_{\rm vir}^2 f_{\rm c} n_{\rm halo}$ of LLSs arise from
halos in the specified mass interval, where $n_{\rm halo}$ is the
volume density of dark matter halos in the selected mass range.  To
account for the remaining systems 
required to give the correct cosmic average line density of LLSs, i.e.
$\ell(z)_{\rm obs}-\ell(z)_{\rm halo}$, we
simply add a random population of absorbers that are not clustered 
to dark matter halos and hence to galaxies. 
In all that follows, we take the incidence to be $\ell(z)_{\rm obs} \equiv 1.5$, 
which is consistent with the observed value for
$N_{\rm HI} \ge 10^{17.2}~\rm cm^{-2}$ LLSs at $z\sim2$ from
\citet{ome13}.

In other words, this model assumes a random, i.e. non-clustered, background of LLSs and of 
a second populations of LLSs that are clustered to galaxies in a selected
mass range. Clearly, this is a rather simplistic approach as, for instance, 
simulations suggest that LLSs are typically clustered to galaxies of different masses
\citep[e.g.][]{koh07,alt11,rah13}. However, this approximation is meant to describe 
the limit in which a subset of LLSs arise from either  low mass galaxies that have 
a small bias compared to the halos here considered, or to a case in which a fraction 
of LLS absorption arises along filaments in the intergalactic medium (IGM) 
and instead traces the Lyman-$\alpha$ 
forest which has a very weak clustering compared to massive halos \citep{mcd03}.
Albeit crude in its treatment of the
baryon distribution around galaxies, this model accurately reproduces the spatial
clustering on large scales that is imposed by structure formation.

For the analysis, we sample these mock universes with random sightlines and we 
compute the projected galaxy-LLS cross-correlation function as described in Equation 
(\ref{binnedchi}) within a velocity window of $\Delta v = \pm 400~\rm km~s^{-1}$,
corresponding to a redshift interval of $\Delta z_0 = 0.008$ or a depth of $12~\rm cMpc$
along the line of sight. This velocity window is suitable for comparisons with observations
as it is large enough to encompass the majority of the denser gas 
($n_{\rm H} \ge 0.1~\rm cm^{-2}$) within 2\rvir\ from the galaxy center, after accounting 
for peculiar velocities along the line of sight. Note that this velocity window is
also larger than typical redshift errors for LBGs  ($\sim 150~\rm km~s^{-1}$).
The resulting $\chi_{\rm gl,\bot}$ from the four different models with varying $f_{\rm c}$ 
are shown in Figure \ref{fig:llslls} (blue lines) as a function of the projected 
separation between galaxies and LLSs. For comparison, we also show the projected two-point 
correlation function of dark matter halos (grey crosses), which we compute by comparing 
the number of galaxy pairs at projected distance $r_\bot$ within the Bolshoi simulation 
to the number of random pairs\footnote{The two-point correlation function
for dark matter halos flattens at scales of $\lesssim 0.8~\rm cMpc/h$ because of halo 
exclusion effects for which two halos cannot occupy the same volume.}. 

Figure \ref{fig:llslls} provides a schematic view of the CGM properties 
that can be extracted from the galaxy-LLS correlation function. First, one can see that 
at projected separations that are typical for the one-halo term
($\sim 0.3-0.4~\rm cMpc/h \sim 2$\rvir), the projected correlation function is
proportional to the covering fraction of optically-thick gas inside the dark matter halos. 
By construction, our models do not incorporate any radial dependence for $f_{\rm c}$
within $r_{\rm \bot} < 2$\rvir. However, our zoom-in simulations exhibit only a shallow radial 
profile for the covering fraction (see e.g. Figure \ref{fig:qsocf}), and 
a modest radial dependence for the projected correlation function up to $\sim 2R_{\rm vir}$
becomes a general prediction. If we adopted a power law form for 
the correlation function $\xi_{\rm gl}(r) \sim (r/r_0)^{-\gamma}$, and fitted only data
interior to $r_{\rm \bot} < 2$\rvir\ that are dominated by this flat one-halo term,  we 
would infer a large correlation length $r_0$ or, equivalently, a shallow exponent 
$\gamma$. A quantitative comparison between the observed and 
predicted galaxy-LLS cross-correlation function offers therefore 
an additional test for theories of gas accretion around galaxies. 

The second feature that is visible in Figure \ref{fig:llslls} is that, 
around $\sim 0.5~\rm cMpc/h$, the projected cross-correlation function exhibits 
a break at the transition between the one-halo term and the two-halo term. 
This feature offers a natural way to define the typical extent of the
CGM in the galaxy population under examination. Finally, at larger
projected separations ($r_\bot \gtrsim 1~\rm cMpc/h$), the two-halo
term of the cross-correlation function traces the (halo mass
dependent) two-point correlation function of the dark matter halos
that host LLSs.  For models with large covering fractions such that
$\ell(z)_{\rm obs} \sim \ell(z)_{\rm halo}$, the galaxy-LLS and halo
correlation functions overlap, while for models with lower $f_{\rm c}$, 
the amplitude of the galaxy-LLS correlation function is
suppressed compared to the halo correlation function due to the
increasingly higher contribution from the background, which in this
particular modelization is randomly distributed, and hence dilutes the
clustering signal.  Note however that the shape of the cross-correlation 
function is preserved for the case of a large random background.

Finally, Figure \ref{fig:llslls} reveals that even models with modest
covering fraction of optically-thick gas as predicted by our zoom-in
simulations exhibit a high amplitude for the projected correlation
function. This is a direct consequence of the limited number of LLSs
that are expected at random within a velocity window of $\Delta v= \pm
400~\rm km~s^{-1}$ from a galaxy. Given the amplitude of the 
correlation function, for models with $f_{\rm c}=0.25$, samples of
$\sim 30$ galaxies-quasar pairs are needed to detect the one-halo
term of the galaxy-LLS correlation function at $\sim 3\sigma$. 
To place interesting constraints on models, samples with at least 80 
galaxy-LLS sightlines are needed. Twice as much pairs are instead 
required for this measurement for the $f_{\rm c}=0.10$ case.
The galaxy-LLS cross-correlation function has already been measured 
on small scales for the quasar host galaxies \citep{hen07,pro13b}. 
However, building up the required statistics to make a
measurement of comparable precision of the LBG-LLS cross-correlation
is a more challenging task. While one can attempt to detect a signal with 
current data, samples that are 5 to 10 times larger than what currently available 
are needed to precisely characterize the distribution of optically-thick gas around 
galaxies.

\subsection{The LLS auto-correlation function}\label{llslls}

To circumvent the observational challenges of building up large
foreground galaxy-background quasar samples, we propose that one
measures the \emph{auto-correlation} function of LLSs, using the 
large existing samples of close quasar pairs, with $\sim 300$ 
pairs currently known at $r_\perp < 200~\rm kpc$ \citep{hen04,hen06a,hen10}.
Further, one can exploit samples of lensed quasars to extend this measurement 
to even smaller scales of $\lesssim 10~\rm kpc$ \citep[e.g.][]{ina12}.
The key advantage of this technique is that LLSs are easy to identify even in 
modest signal-to-noise spectra, and hence large samples of LLS pairs can be assembled 
at $z\sim 2-3$.

The idea of the LLS auto-correlation function builds on previous work 
that has shown the power of correlating absorption systems along multiple quasar sightlines
to reveal the spatial distribution of hydrogen or metals in the IGM 
\citep[e.g.][]{mcd03,mar10,slo11,fon12}. 
We now generalize the formalism presented for the galaxy-LLS correlation function 
to the case of two intervening LLSs (i.e. systems that are not 
physically associated to the background quasars) in the foreground of 
quasar pairs with projected separation $r_{12}$.
Next, we will show using numerical models that the LLS auto-correlation function contains 
the same information about the CGM of galaxies as is encoded in the galaxy-LLS 
cross-correlation function. Thus, if LLSs are associated to galaxies, searches for 
LLSs in quasar pairs provide a powerful statistical way to characterize the CGM in 
high redshift galaxies, without the need to identify individual galaxy-LLS associations.

\subsubsection{Formalism}

The formalism to compute the LLS auto-correlation function closely follows the approach 
used to compute the galaxy-LLS cross-correlation function. For a random quasar sightline, 
the probability to find a LLS is $P_1=\ell(z)\Delta z_1$, where $\Delta z_1$ is the useful 
redshift path that can be searched for absorption lines. Once a LLS is found at redshift 
$z_{\rm lls,1}$ the probability of finding a second LLS within $\pm \Delta v$ 
from the redshift of the first LLS along a second sightline at distance $r_{\rm 12}$ is 
\begin{equation}
F_{\rm C} (\Delta z'_0,r_{\rm 12}) = \ell (z_{\rm lls,1}) [1+\chi_{\rm LL,\bot} (\Delta z'_{\rm 0}, r_{\rm 12})] \Delta z'_0\:,
\end{equation}
where $\Delta z'_{\rm 0} = 2 \Delta v(1+z_{\rm lls, 1})/c$ and 
$\chi_{\rm LL,\bot}(\Delta z'_{\rm 0}, r_{\rm 12})$ expresses the 
projected LLS auto-correlation function. As we will show in the following, if LLSs 
mostly arise from galaxy halos, $F_{\rm C}$ is directly related to the covering fraction 
$f_{\rm c}$ of optically-thick gas in the CGM in the galaxy population from which LLSs arise.

As previously done for the galaxy-LLS correlation 
function, we can relate $\chi_{\rm LL,\bot}$ to the LLS auto-correlation 
function $\xi_{\rm LL}(r)$ in real space following Equation (\ref{intchi}).
Altogether, assuming $\ell (z_{\rm lls,2}) \approx \ell (z_{\rm lls,1})$,
the probability to find a pair of LLSs in the foreground of a quasar pair becomes  
\begin{equation}
P_2 (\Delta z_1,\Delta z'_0,r_{\rm 12}) \approx \ell^2 (z_{\rm lls,1})
[1+\chi_{\rm LL,\bot} (\Delta z'_{\rm 0}, r_{\rm 12})] \Delta z'_0\Delta z_1\:.
\end{equation}
 
For an ensemble of quasar pairs, one can measure the 
projected LLS auto-correlation function $\chi_{\rm LL,\bot}$ in bins of $r_{12}$ 
following Equation \ref{binnedchi}.

\subsubsection{Numerical models}

Provided that the population of LLSs can be identified with the CGM of galaxies 
\citep{koh07,fum11,van12,fum13}, a LLS detected at redshift $z_{\rm lls,1}$ along one sightline
signals the presence of a galaxy, which lie at an unknown projected distance $r_{\rm lg}$. 
Therefore, even without identifying the galaxies that are responsible for the absorption, one 
can use a second sightline at projected separation $r_{12}$ from the first quasar to probe 
the distribution of optically-thick gas in the galaxy halo. The LLS auto-correlation function 
is thus analogous to the galaxy-LLS correlation function, providing a statistical way of 
mapping the CGM of distant halos without explicitly identifying galaxy-LLS associations.

To illustrate this point with numerical models, we generate a new realization from the 
Bolshoi simulation assuming $f_{\rm c}=0.25$, such that the majority of LLSs arise from 
halos with masses $5\times 10^{11}-5\times 10^{12}~\rm M_\odot$. 
We then sample the simulated box with pairs of sightlines with
separations $r_{12}$ and compute the projected LLS auto-correlation
function as described in Equation (\ref{binnedchi}), that is by
comparing the pairs of LLSs with a given $r_{12}$ and within a
velocity window of $\Delta v = \pm 400~\rm km~s^{-1}$ to the random
expectation.  The resulting LLS auto-correlation function is shown
with a red dashed line in Figure \ref{fig:llslls}.  Note that in this
figure the projected separation on the x-axis corresponds to the
distance between quasar pairs for the LLS auto-correlation function,
while it corresponds to the separation between a quasar sightline and
a galaxy (assumed to be at the center of the dark matter halo in our
models) for the galaxy-LLS correlation function.

As is evident from Figure \ref{fig:llslls}, the projected LLS auto-correlation function 
closely resembles the projected galaxy-LLS correlation function for 
$f_{\rm c}=0.25$. The only difference
is that for the galaxy-LLS pairs,  the galaxy is always at the center of the dark
matter halo, whereas for the LLS-LLS pairs, the halo centers are offset by a random amount
relative to the two quasars probing the LLSs, and thus  
$\chi_{\rm LL,\bot}(\Delta z'_{\rm 0}, r_{12})$
reflects the properties of the halo gas smoothed on scales that are comparable to the
size of the CGM, or $\sim 2R_{\rm vir}$ in our numerical models. Nevertheless, 
the LLS auto-correlation function encodes all the information we previously discussed 
for the galaxy-LLS correlation function. This includes a flat one-halo term
with an amplitude that varies with the covering fraction in 
the host halos, a one-halo to two-halo term transition that 
can be used to define the characteristic size of the CGM in the galaxies 
where LLSs arise, and a two-halo term which traces the large-scale clustering of the 
underlying dark matter halos hosting LLSs. The fact that the large 
scale LLS correlation traces the clustering of dark matter halos is a 
trivial consequence of how we constructed our models by associating LLSs only to 
dark matter halos in a selected mass range. However, this exercise shows 
how measurements of the large-scale LLS auto-correlation function can be used
to determine the typical halo masses which host LLSs, 
a key unknown quantity that currently hampers the interpretation of the observed LLS 
properties \citep[e.g.][]{fum13}.

In Figure  \ref{fig:llslls}, we also show the LLS auto-correlation function in a realization 
with $f_{\rm c}=0.10$ (red dotted line)  and $f_{\rm c}=0.05$ (red dashed triple-dotted line).
In the latter case, the majority of LLSs ($\sim 80\%$) are not clustered 
to galaxies, but they reside in a random background. 
As expected, one can see how the projected auto-correlation function approaches zero.
A comparison between the three models with $f_{\rm c}=0.05,0.10,0.25$ is useful to 
highlight the two extreme behaviors that the LLS auto-correlation function may reflect. 
For high covering fractions, or more generally when the product of the covering 
fraction and the size of the CGM is large (as suggested by current observations), 
the number of LLSs that are associated to galaxies exceeds the number of 
LLSs in a random (non clustered) background. In this case, $\chi_{\rm LL,\bot}\gg 0$ and thus 
the LLS auto-correlation function yields information on the CGM properties. 
Conversely, if either $f_{\rm c}$ is small or the radial profile of optically-thick 
gas in the CGM is very steep, then the number of LLSs associated to galaxies is 
much smaller than the number of LLSs in a random background and
$\chi_{\rm LL,\bot}\sim 0$.  In this case, a measurement of the LLS auto-correlation 
function can be used to conclude that LLSs are not associated to massive galaxies, 
but rather they originate from a more weakly clustered population, e.g. the 
Lyman-$\alpha$ forest. It should also be noted that if the fraction of LLSs that 
are associated to galaxies evolves with redshift \citep[cf.][]{fum13}, then  
the auto-correlation function of LLSs will evolve accordingly.

From the above discussion, it follows the LLS auto-correlation function encodes information 
on the cross section of optically thick gas around galaxies, similarly to the 
measurement of the cross-correlation function of damped Lyman-$\alpha$ systems 
with the Lyman-$\alpha$ forest \citep[see][]{fon12}. For instance, 
in constructing these simple models, we have assume a mass-independent covering 
fraction $f_{\rm c}$ within 2\rvir, which implies a mass dependent cross section 
$\sigma_{\rm lls} (M) \propto M_{\rm vir}^{2/3}$. In computing the LLS auto-correlation 
function,  $\sigma_{\rm lls}$ determines the weight with which each halo of a given mass 
contributes to the observed value of $\chi_{\rm LL}$. For this reason, a precise measurement 
of the auto-correlation of optically-thick systems around quasar pairs provides 
a way to constrain the mass-dependent cross section 
$\sigma_{\rm lls} (M) \propto M_{\rm vir}^\alpha$, a quantity for which theoretical 
predictions exist from hydrodynamic simulations \citep[e.g.][]{bir13}.

We conclude by noting that, thanks to the large samples of quasar spectra 
that currently are or will be soon available from surveys like the Sloan Digital Sky 
Survey \citep[SDSS; e.g.][]{par12}, a measurement of the LLS auto-correlation function 
can be obtained at large scales ($\gtrsim 1 h^{-1}\,{\rm Mpc}$) at 
redshifts $z\ge 3$. The minimum angular separation is set by the fiber collision 
limit in the spectroscopic survey, which severely limits the number of close quasar pairs
with available spectroscopy, while the redshift constraint is currently set by the
throughput of the survey spectrograph. However, because of the SDSS color-selection bias 
that preferentially selects quasars with LLS absorption at $z<3.6$ \citep{wor11,fum13}, 
additional investigation is needed to establish whether the redshift limit has to be 
restricted to $z\ge 3.6$.

To measure the LLS auto-correlation function on small scales, hundreds of spectroscopically 
confirmed quasar pairs with projected separations between 
$30\,h^{-1}\,{\rm kpc}-1 h^{-1}\,{\rm Mpc}$ have now been 
discovered via follow-up spectroscopy of the SDSS imaging \citep{hen04,hen06a,hen10}. 
This sample allows a precise measurement of the small-scale clustering at high confidence 
level ($>5\sigma$ for $f_{\rm c}=0.25$). Although the majority of useful pairs are currently 
between $z\sim 2-3$ requiring the use of space-based facilities to identify LLSs in the 
foreground of these quasar pairs, a precise measurement of the LLS auto-correlation function on 
all scales can potentially be achieved in the near future. 

\section{Summary and conclusions}\label{concl}

We have analyzed the hydrogen distribution in the surroundings of 21 galaxies at $z\sim2-3$ 
that have been simulated at high resolution with virial masses 
$M_{\rm vir} \sim 2\times 10^{11} - 4\times 10^{12}~\rm M_\odot$. 
After post-processing these simulations with a Monte Carlo radiative transfer code
to identify regions that retain enough neutral hydrogen to remain optically thick to Lyman 
continuum radiation, we have directly compared the covering fraction of optically-thick 
gas in simulations and observations of $z\sim 2$ LBGs and quasar host galaxies. We have also 
presented a formalism to compute the galaxy-LLS cross-correlation function and the LLS 
auto-correlation function, and we have provided simple estimates for these quantities 
using numerical simulations. Our main findings can be summarized as follows. 

The covering fractions of optically-thick gas within the virial radius
of the simulated galaxies range between $f_{\rm c} \sim 0.05-0.2$,
where the large scatter is driven by intrinsic variation in the gas
distributions around individual halos. Within 2\rvir, we have found
instead $f_{\rm c} \sim 0.01 - 0.13$, implying that the area subtended
by optically-thick gas within \rvir\ and between $R_{\rm vir} < R <
2R_{\rm vir}$ is approximately the same.  While our simulations
exhibit the expected increase in the average gas temperature and an increase
in the mass fraction of hot gas above virial masses for which stable
virial shocks form, we have found that the mass fraction of cold gas with
$T< 3\times 10^4~\rm K$ is only weakly dependent on halo mass.
Further, at $z\ge 2$, we have not found any strong dependence of the covering fraction on 
the halo mass, even beyond the critical mass for the formation of virial shocks.

Once compared to observations of 10 galaxy-quasar pairs at $z\sim 2-3$,
these simulations are statistically consistent with the observed covering fraction 
of optically-thick gas inside the virial radius. However, current samples are too
small to make a conclusive comparison, preventing stringent tests for current 
theories of cold gas accretion. Conversely, simulated halos at 
$M_{\rm vir} \ge 10^{12}~\rm M_\odot$ exhibit covering 
fractions at all radii that significantly underestimate the values observed in the
surroundings of quasar host galaxies. This discrepancy reveals that
our numerical models do not fully capture all of the physical
processes necessary to describe the gas distribution around massive
halos. At present, we do not know the explanation for this disagreement, but
issues that should be investigrated in future work are i)
modeling the effects of stronger (AGN) feedback and/or small-scales hydrodynamic 
instabilities than what is currently implemented in our simulations, ii) or a better 
understanding on how the properties of quasar host galaxies, in particular their
star-formation rates or gas masses, compare to other
populations of star-forming galaxies such as the LBGs. 

Further, we have showed how mesurements of the galaxy-LLS correlation
function can be used to measure the ccovering fraction of LLSs around
galaxies. The flat radial depence of the covering fraction interior to
the $R_{\rm vir}$ predicted by our simulations, implies that the
projected galaxy-LLS correlation function will exhibit a shallow
radial dependence on small-scales that probe the one-halo term. We
have also showed that the transition between the one-halo term and two-halo
term imprints a feature in the projected cross-correlation function
that can be used to define the spatial extent of the CGM. 

Finally, under the assumption that LLSs are statistically associated
to galaxy halos of a given mass range, we have proposed a measurement of
the LLS auto-correlation function using quasar pair sightlines, to map
the spatial distribution of optically-thick gas around galaxies,
without the need to identify individual galaxy-LLS associations.  Our
numerical models show that the LLS auto-correlation function encodes
the same information contained in the galaxy-LLS correlation function
(both covering fraction of optically-thick gas, the characteristic
size for the CGM), but
smoothed on scales comparable to the typical size of the CGM. 
Furthermore, we have highlighted that at large separations the two-halo term of the 
LLS auto-correlation function traces the two-point correlation
function of the dark matter halos hosting LLSs,  providing long-sought 
information about the typical mass of the halos that host LLSs.

While our analysis underscores a still incomplete view of the gas distribution around massive
galaxies, in this paper we have outlined a possible path towards an improved knowledge 
of the properties of the halo gas in the distant Universe. 
In the long term, the increasing availability of samples of quasar-galaxy 
pairs will offer a direct way to map the radial distribution of optically-thick gas at 
high redshift. Measurements of the galaxy-LLS correlation can be
compared to different sets of simulations, providing additional insights into the 
processes that regulate the structure of the CGM and ultimately the formation and 
evolution of galaxies. Given the current availability of large spectroscopic samples of quasars 
and hundreds of quasar pairs with small projected separations, it is also 
possible to compute the LLS auto-correlation function to obtain the first 
view of the spatial distribution of optically-thick gas in the high-redshift Universe.

As discussed, this measurement would provide an important test for the 
cold-stream paradigm, as well as a solid empirical assessment of whether LLSs 
arise primarily in the CGM of galaxies at $z \sim 2-3$. Provided that the connection between 
LLSs and halo gas can be robustly established, analysis 
of the physical properties of these absorbers would then offer a powerful way to 
map the metal distribution in proximity to galaxies in the distant Universe. 
Further, better knowledge on the clustering of LLSs would affect estimates
for the extragalactic UV background, that depend strongly on the distribution of 
optically-thick gas. It is therefore clear that an improved understanding of how LLSs cluster 
around galaxies and around themselves would constitute an important step forward that will 
impact several areas of study.

\acknowledgements

The simulations were performed at NASA Advanced
Supercomputing (NAS) at NASA Ames Research
Center, at the National Energy Research Scientific
Computing Center (NERSC) at Lawrence Berkeley
Laboratory, and in the astro cluster at The Hebrew University.
We acknowledge useful conversations with Andrew Benson and we thank 
Claude-Andr\'e Faucher-Gigu\`ere for his comments on this manuscript.
We also thank the referee for suggestions that have improved this paper. 
Support for this work was provided by NASA to MF through Hubble Fellowship 
grant HF-51305.01-A awarded by the Space Telescope Science Institute, 
which is operated by the Association of Universities for Research in 
Astronomy, Inc., for NASA, under contract NAS 5-26555. 
MF thanks the members of CCAPP and the astronomy department at Ohio 
State University for their hospitality during a visit made possible by the Price Prize 
and for interesting discussions on the LLS correlation function.
JXP is supported by NSF grant AST-1010004. AD acknowledges support by ISF grant 24/12, 
by GIF grant G-1052-104.7/2009, by a DIP-DFG grant. AD and JP acknowledge support by NSF 
grant AST-1010033. DC is supported by the JdC subprogramme JCI-2010-07122.

\appendix

\section{Method for Calculating Ionizing Radiation Transport}\label{ratraapp}

To calculate the transport of ionizing radiation, we used a 3D Monte Carlo (MC) code, 
derived from the SEDONA code framework \citep{kas06} and using an approach similar to that 
described in \cite{woo00}.  The radiation field is represented by a large number 
($N_{\rm p} \sim 10^8$) of discrete photon packets, which are propagated throughout 
absorption or scattering events until they escape the simulation domain.  This approach 
permits an arbitrary number of individual sources and conserves energy by construction.  
In addition, we are able to properly model the diffuse ionizing  radiation field arising 
from recombinations to the ground state -- i.e., we do not make the ``on-the-spot'' 
approximation.  The main disadvantage of MC methods is that a large number of photon packets 
must be used to overcome statistical noise, however the method scales well and can be run 
on massively parallel machines.

The calculation in this paper ignores time-dependent effects and makes the assumption of 
ionization equilibrium \citep[cf.][]{can11,opp13}.  Each photon packet carries a 
``luminosity'' of $L_{\rm p} = (L_{\rm ubv} + L_{\rm loc})/N_{\rm p}$, where $L_{\rm ubv}$ and 
$L_{\rm loc}$ are the total ionizing luminosities from the UVB and from local sources, 
respectively.  A fraction $f = L_{\rm ubv}/(L_{\rm ubv} + L_{\rm loc})$ of the photon packets 
are selected to represent the UVB. These packets are initially distributed randomly over 
the surface of a sphere of radius $R_{\rm ubv}$, chosen to be larger than the simulation box.  
The effective luminosity of the UVB from this outer surface is given by
\begin{equation}
L_{\rm ubv} =  4 \pi R_{\rm ubv}^2  \int d\nu~ \Jbk(\nu),
\end{equation}
where we choose a UVB mean intensity at $z \sim 2$ and $z\sim 3$ (according to the
redshift of each snapshot) from \citet{haa12}. For simplicity, the spectrum of the UVB is 
assumed to be flat over the energy range $1 - 4$ Rydberg ($238-912$~\AA) corresponding to 
the wavelength region from the HeII ionization threshold to the hydrogen 
threshold.  

To produce a homogenous UVB radiation field within the simulation domain, the 
directionality of the packets is sampled from a $\cos \theta$ angular distribution,  
where $\theta$ is the angle between the packet direction vector and the inward local radial 
vector.  To test the validity of this approach, we ran a calculation using $N_p  = 10^8$ 
packets and assuming that the opacity throughout the domain was zero everywhere.  
A nearly uniform UVB of the desired mean intensity was achieved in the simulation box, 
with random errors of the order of $1\%$.
 
The remaining $1-f$ of the packets are selected to represent the ionizing radiation from 
local sources, and are emitted isotropically from locations given by the star particles 
from the hydrodynamic simulation.  The probability of a packet being emitted from a 
given star particle is proportional to the UV luminosity of that star particle
\citep[see][]{fum11}. As with the UVB, the spectrum of local sources is assumed to be
flat between 1 and 4 Rydberg.

Packets are propagated through the AMR grid until they are absorbed or escape the domain.   
The mean free path to a photoionization interaction with neutral hydrogen is given by 
$1/n_{\rm HI} \sigma_p(\nu)$ where $\sigma_p(\nu)$ is the photoionization cross-section.  
Whenever a photon ionizes a hydrogen atom, the atom is assumed to recombine either  
to the ground state or to an excited state followed by a cascade to the ground state.  
The former case corresponds to an ``effective scattering'' in which an ionizing photon 
is immediately re-emitted in a new direction.   The probability of this occurring is 
$P_s = \alpha_1/\alpha_A$ where $\alpha_1$ is the recombination coefficient to the 
ground state while $\alpha_A$ is the coefficient for recombination to all levels 
including the ground state.    We assume a constant value $P_s = 0.38$ appropriate 
for gas at $T = 10^4$~K, as this quantity is only weakly dependent on temperature.  
At each ionization interaction event, a random number is chosen to determine whether 
recombination to the ground state occurs; if so, the photon packet is redirected 
isotropically, assigned a new wavelength from the local emissivity function, and 
its propagation continues until a true absorption occurs.

To calculate the state of the gas, ionization equilibrium is assumed,
\begin{equation}
R_{\rm pi} + R_{\rm ci} = R_{\rm rr},
\label{Eq:ion_eq}
\end{equation}
where $R_{\rm pi}$ and $R_{\rm ci}$ are, respectively, the hydrogen photoionization and 
collisional ionization rates, and $R_{\rm rr}$ is the hydrogen radiative recombination 
rate (all per cm$^3$).   At these low densities collisional recombination can be ignored. 
The photoionization rate is given by
\begin{equation}
R_{\rm pi} = 4 \pi  n_{\rm H}  x_{\rm HI}  \int d\nu~ \frac{J_\nu(\nu) }{h \nu}  \sigma_p(\nu)
\label{Eq:photoion}
\end{equation}
where $n_{\rm H}$ is the total hydrogen density, $x_{\rm HI}$ is the fraction of hydrogen in 
the neutral state, and  $J_\nu$ is the mean intensity of the ionizing radiation field.
The radiative recombination rate is
\begin{equation}
R_{\rm rr} = n_e n_p \alpha_{A} 
\end{equation}
where $n_p$ is the proton density, and $n_e$ the electron density.   For these calculations,
we assume helium is neutral and does not contribute to the free electron density. In that 
case, we take $n_e = n_p = (1 - x_{\rm HI}) n_{\rm H}$.  Expressions for $\sigma_p$ and 
$\alpha_{A}(T)$ are taken from \citet{ver96} and \citet{ver96b}, respectively, 
and the collisional ionization rate from \citet{jef68}.

During the MC procedure, an estimate of the photoionization rate (Equation (\ref{Eq:photoion})) 
is constructed in each cell by tallying all traversing packets 
\citep[e.g.,][]{luc02,woo00}
\begin{equation}
R_{\rm pi}/x_{\rm HI} = \frac{n_{\rm H}}{V} \sum_i \frac{L_p \sigma_p(\nu)}{h \nu} l_i
\label{Eq:estimate}
\end{equation}
where $V$ is the cell volume, and the sum runs over all steps of length $l_i$ that occur for 
packets passing through the cell.

An iterative approach is used to converge the model to ionization equilibrium.  Initially, a 
guess is made as to the neutral fraction $x_{\rm HI}$ in all cells.  We then follow the MC 
transport and construct the estimator Equation (\ref{Eq:estimate}).  By solving Equation 
(\ref{Eq:ion_eq}), we obtain a new value for the neutral fraction in each cell.   The MC 
transport routine is then rerun, and a new estimate of $x_{\rm HI}$ is derived.  This procedure 
is iterated until the ionization state no longer changes significantly from one iteration 
to the next.   To speed convergence, we adopt as an initial guess that hydrogen is 
completely ionized everywhere, as in this case photons packets can propagate information 
across the entire domain.  We find that 12 iterations are sufficient for convergence.  

\begin{figure}
\centering
\includegraphics[angle=90,scale=0.31]{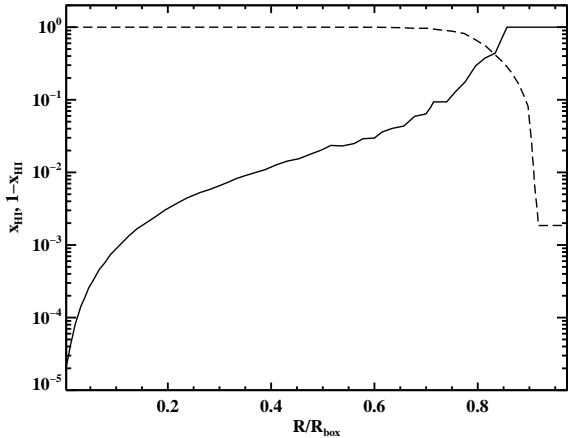}
\caption{Results for test 1 of \citet{ili06} used to validate the photoionization calculation.  The problem consists of a homogenous distribution of gas of number density $n_H = 10^{-3}~\cc$ with a central source of ionizing luminosity  $Q = 5 \times 10^{48}~{\rm s^{-1}}$.  The solid line shows the neutral fraction and the dashed line the ionized fraction of hydrogen, plotted as a function of distance from the center.  The results can be compared the calculations shown in the right hand panel of figure 8 of \citet{ili06}.
\label{Fig:uv_test}}
 \end{figure}

To validate the photoionization code, we perform test 1 of \citet{ili06} which consists
of a box of dimension $R_{\rm box}=6.6$~kpc with uniform gas number density $n_H = 10^{-3}~\cc$. ú 
A source of  ionizing photons with a production rate $Q = 5 \times 10^{48}~{\rm s^{-1}}$
is placed at the center of the box. Figure~\ref{Fig:uv_test} shows the resulting equilibrium 
ionization structure, which is in agreement with the converged structure presented in 
figure~8 of \citet{ili06}.

\end{document}

%% file: table_prop.tex
\begin{deluxetable*}{lccccccccc}
\tablewidth{0pc}
\tablecaption{Properties of the simulated galaxies included in this study.}
\tablehead{\colhead{Model}
& \colhead{Exp. Factor} 
& \colhead{Redshift} 
& \colhead{$R_{\rm vir}$} 
& \colhead{$M_{\rm vir}$} 
& \colhead{$M_{\rm dm}$} 
& \colhead{$M_{\rm star}$} 
& \colhead{$M_{\rm gas}$} 
& \colhead{SFR} 
& \colhead{$\phi_{\rm cold}$} \\
& & & (kpc) & ($10^{12}$ M$_\odot$) & ($10^{12}$ M$_\odot$) &  ($10^{11}$ M$_\odot$) &   ($10^{11}$ M$_\odot$) 
& (M$_\odot$ yr$^{-1}$) & ($T<3\times 10^4~\rm K$)}
\startdata
\multicolumn{10}{c}{$z\sim 3$ sample} \\
MW1 &	 0.24&  3.17&  29.8&  0.05&  0.05&  0.03&  0.03&   4.4 & 0.46 \\
MW2 &	 0.24&  3.17&  57.0&  0.31&  0.20&  0.75&  0.35& 148.2 & 0.80 \\
MW3 &	 0.24&  3.17&  38.0&  0.11&  0.09&  0.09&  0.05&  14.4 & 0.46 \\
MW4 &	 0.24&  3.17&  51.8&  0.27&  0.23&  0.28&  0.13&  15.8 & 0.52 \\
MW7 &	 0.25&  3.00&  49.5&  0.21&  0.18&  0.25&  0.07&  18.3 & 0.42 \\
MW8 &	 0.25&  3.00&  42.5&  0.13&  0.11&  0.14&  0.07&   5.5 & 0.61 \\
MW9 &	 0.25&  3.00&  37.8&  0.10&  0.08&  0.11&  0.04&  11.2 & 0.54 \\
MW10&    0.25&  3.00&  38.8&  0.10&  0.09&  0.10&  0.04&   9.7 & 0.39 \\
MW11&    0.25&  3.00&  40.0&  0.11&  0.10&  0.13&  0.03&   6.2 & 0.46 \\
MW12&    0.25&  3.00&  80.0&  0.87&  0.74&  0.97&  0.42&  59.6 & 0.38 \\
SFG1&    0.25&  3.00&  91.2&  1.34&  1.14&  1.48&  0.47&  64.1 & 0.26 \\
SFG4&    0.25&  3.00&  60.8&  0.39&  0.33&  0.42&  0.19&  55.0 & 0.35 \\
SFG5&    0.25&  3.00&  66.0&  0.51&  0.44&  0.52&  0.20&  28.7 & 0.46 \\
SFG7&    0.25&  3.08& 102.0&  1.90&  1.60&  1.93&  1.08& 128.5 & 0.51 \\
SFG8&    0.25&  3.00&  82.8&  1.00&  0.85&  1.20&  0.35&  53.0 & 0.36 \\
SFG9&    0.25&  3.00&  88.5&  1.21&  1.02&  1.47&  0.46&  64.6 & 0.34 \\
VL01&    0.25&  3.00&  74.5&  0.73&  0.62&  0.90&  0.23&  53.5 & 0.39 \\
VL04&    0.25&  3.00&  77.0&  0.80&  0.68&  0.93&  0.31&  32.3 & 0.33 \\
VL06&    0.25&  3.00&  61.8&  0.41&  0.35&  0.49&  0.13&   6.6 & 0.40 \\
VL09&    0.25&  3.00&  53.2&  0.26&  0.23&  0.27&  0.12&  31.7 & 0.49 \\
VL11&    0.25&  3.00&  63.2&  0.45&  0.38&  0.54&  0.17&  41.8 & 0.28 \\
\multicolumn{10}{c}{$z\sim 2$ sample} \\
MW1 &	 0.34&  1.94& 106.8&  0.88&  0.75&  0.82&  0.46&  62.0 &  0.40     \\
MW2 &	 0.34&  1.94& 108.8&  0.91&  0.54&  2.75&  0.94& 187.9 &  0.72     \\
MW3 &	 0.34&  1.94& 104.0&  0.81&  0.69&  0.71&  0.42&  71.5 &  0.43     \\
MW4 &	 0.34&  1.94& 129.0&  1.52&  1.29&  1.54&  0.76&  77.2 &  0.31     \\
MW7 &	 0.33&  2.03&  73.0&  0.31&  0.25&  0.46&  0.10&  17.3 &  0.29     \\
MW8 &	 0.33&  2.03&  70.5&  0.27&  0.23&  0.27&  0.11&   5.4 &  0.42     \\
MW9 &	 0.33&  2.03&  58.2&  0.15&  0.13&  0.19&  0.06&   2.0 &  0.48     \\
MW10&    0.33&  2.03&  99.0&  0.77&  0.66&  0.71&  0.31&  24.1 &  0.44     \\
MW11&    0.33&  2.03&  87.2&  0.52&  0.45&  0.50&  0.21&  32.5 &  0.33     \\
MW12&    0.33&  2.03& 127.8&  1.64&  1.36&  2.01&  0.78&  31.9 &  0.26     \\
SFG1&    0.33&  2.03& 126.8&  1.61&  1.34&  2.06&  0.65&  21.3 &  0.24     \\
SFG4&    0.33&  2.03& 110.8&  1.06&  0.90&  1.15&  0.50&  19.4 &  0.29     \\
SFG5&    0.33&  2.03& 122.0&  1.35&  1.14&  1.50&  0.60&  31.7 &  0.33     \\
SFG7&    0.28&  2.51& 152.5&  4.18&  3.55&  4.10&  2.18& 266.3 &  0.27     \\
SFG8&    0.33&  2.03& 119.8&  1.36&  1.13&  1.68&  0.55&  45.7 &  0.15     \\
SFG9&    0.33&  2.02& 133.5&  1.85&  1.51&  2.43&  0.93&  29.9 &  0.24     \\
VL01&    0.33&  2.03& 115.0&  1.17&  0.97&  1.51&  0.57&  49.8 &  0.38     \\
VL04&    0.33&  2.03& 107.8&  0.99&  0.82&  1.31&  0.39&  26.6 &  0.17     \\
VL06&    0.33&  2.03&  98.2&  0.74&  0.62&  0.93&  0.25&   9.8 &  0.29     \\
VL09&    0.33&  2.03&  83.8&  0.46&  0.39&  0.56&  0.18&  10.2 &  0.35     \\
VL11&    0.33&  2.02& 129.2&  1.69&  1.43&  2.01&  0.62& 109.5 &  0.19
\enddata\label{tab:galprop}
\end{deluxetable*}